\documentclass[aps,preprint,groupedaddress,showpacs]{revtex4}

\begin{document}
\title{Derivation via free energy conservation constraints of 
        gyrofluid equations with finite-gyroradius electromagnetic
        nonlinearities} 
\author{B. Scott}
\email[email: ]{bds@ipp.mpg.de}
\homepage[\\ URL: ]{http://www.rzg.mpg.de/~bds/}
\affiliation{Max-Planck-Institut f\"ur Plasmaphysik, 
		Euratom Association,
                D-85748 Garching, Germany}

\date{Mar 2010}

\begin{abstract}
The derivation of electromagnetic gyrofluid equations is made systematic
by using the Hermite polynomial form of the underlying delta-f
gyrokinetic distribution function.  The gyrokinetic free-energy
functional is explicitly used to set up the model.
The gyrofluid free energy follows directly.
The interaction term in the gyrokinetic Lagrangian
is used to obtain the gyrofluid counterpart, from which the polarisation
equation follows.  One closure rule is decided for taking moments over
the kinetic gyroaveraging operator.  These steps fix the rest of the
derivation of the conservative part of the gyrofluid equations.
Dissipation is then added in a form to obtain positive definite
dissipation and to obtain the collisional fluid equations in their
appropriate limit.  Existing results are recovered, with the addition of
a completely consistent model for finite gyroradius effects in the
nonlinearities responsible for magnetic reconnection.
\end{abstract}


\maketitle

\def\emskip{\hskip 1 em}
\def\hfb{\hfil\break}
\def\etc{{\it etc.}}
\def\visavis{{\it vis-a-vis}\ }
\def\ie{{\it i.e.}}
\def\eg{{\it e.g.}}
\def\etal{{\it et al}}
\def\ua{u.a.\ }
\def\dh{d.h.\ }
\def\zb{z.B.\ }
\def\bzw{bzw.\ }
\def\usw{usw.\ }

\def\idelta{$i$-delta}


\def\half{ {1\over 2} }
\def\third{ {1\over 3} }
\def\fourth{ {1\over 4} }
\def\tth{ {2\over 3} }
\def\twothirds{ {2\over 3} }
\def\threehalves{ {3\over 2} }
\def\fivehalves{ {5\over 2} }
\def\fivethirds{ {5\over 3} }
\def\sevenhalves{ {7\over 2} }
\def\threeh{\threehalves}
\def\eps{\epsilon}
 
\def\grapprox{\mathop{\lower.5ex \hbox{$\buildrel{\fivesy >}\over{\fivesy\sim}$}} \nolimits}
\def\lsapprox{\mathop{\lower.5ex \hbox{$\buildrel{\fivesy <}\over{\fivesy\sim}$}} \nolimits}
\def\grls{\mathop{\lower.5ex \hbox{$\buildrel{\fivesy >}\over{\fivesy <}$}} \nolimits}

\def\vec#1{{\bf #1}}
\def\tsr#1{{\secfnt #1}}
\def\avg#1{\left\langle #1 \right\rangle}
\def\abs#1{\left\vert #1 \right\vert}
\def\prf#1{\overline{#1}}

\def\max{{}_{{\rm max}}}
\def\min{{}_{{\rm min}}}

\def\minus{\mathop{\hbox{--}}\nolimits}

\def\re{\mathop{\rm Re}\nolimits}
\def\im{\mathop{\rm Im}\nolimits}
\def\csch{\mathop{\rm csch}\nolimits}
\def\sech{\mathop{\rm sech}\nolimits}
\def\diag{\mathop{\rm diag}\nolimits}
\def\Max{\mathop{\rm Max}\nolimits}
\def\Min{\mathop{\rm Min}\nolimits}
\def\nint{\mathop{\rm NINT}\nolimits}
\def\mod{\mathop{\rm mod}\nolimits}
\def\det{\mathop{\rm det}\nolimits}
\def\Tr{\mathop{\rm Tr}\nolimits}
\def\sign{\mathop{\rm sign}\nolimits}

\def\LBR{\left\lbrace}
\def\RBR{\right\rbrace}
\def\LB{\left\lbrack}
\def\RB{\right\rbrack}
\def\LP{\left (}
\def\RP{\right )}
\def\qqquad{\qquad\qquad}
\def\qqqquad{\qquad\qquad\qquad}
\def\Det#1{\left\vert\matrix{#1}\right\vert}

\def\pt{\partial}

\def\pzz#1{{\partial #1\over\partial z}}
\def\pxx#1{{\partial #1\over\partial x}}
\def\pyy#1{{\partial #1\over\partial y}}
\def\pww#1{{\partial #1\over\partial w}}
\def\pss#1{{\partial #1\over\partial s}}
\def\prr#1{{\partial #1\over\partial r}}
\def\prhrh#1{{\partial #1\over\partial \rho}}
\def\pthth#1{{\partial #1\over\partial \theta}}
\def\pchch#1{{\partial #1\over\partial \chi}}
\def\ppsps#1{{\partial #1\over\partial \psi}}
\def\pzeze#1{{\partial #1\over\partial \zeta}}
\def\pphph#1{{\partial #1\over\partial \phi}}
\def\ptt#1{{\partial #1\over\partial t}}
\def\pVV#1{{\partial #1\over\partial V}}
\def\phh#1{{\partial #1\over\partial \theta}}
\def\pvhvh#1{{\partial #1\over\partial \vartheta}}
\def\pxixi#1{{\partial #1\over\partial \xi}}
\def\dtt#1{{d #1\over dt}}
\def\dss#1{{d #1\over ds}}
\def\drr#1{{d #1\over dr}}
\def\pprr#1{{\partial^2 #1\over\partial r^2}}
\def\pprhrh#1{{\partial^2 #1\over\partial \rho^2}}
\def\ppss#1{{\partial^2 #1\over\partial s^2}}
\def\ppxx#1{{\partial^2 #1\over\partial x^2}}
\def\ppxy#1{{\partial^2 #1\over\partial x\partial y}}
\def\ppxs#1{{\partial^2 #1\over\partial x\partial s}}
\def\ppys#1{{\partial^2 #1\over\partial y\partial s}}
\def\ppyy#1{{\partial^2 #1\over\partial y^2}}
\def\ppzz#1{{\partial^2 #1\over\partial z^2}}
\def\pptt#1{{\partial^2 #1\over\partial t^2}}
\def\ppVV#1{{\partial^2 #1\over\partial V^2}}
\def\ppphph#1{{\partial^2 #1\over\partial \phi^2}}
\def\ppthth#1{{\partial^2 #1\over\partial \theta^2}}
\def\pphh#1{{\partial^2 #1\over\partial \theta^2}}
\def\ppvhvh#1{{\partial^2 #1\over\partial \vartheta^2}}
\def\ppxixi#1{{\partial^2 #1\over\partial \xi^2}}
\def\ppzeze#1{{\partial^2 #1\over\partial \zeta^2}}
\def\pphze#1{{\partial^2 #1\over\partial\theta\partial\zeta}}
\def\ppz#1{\partial #1/\partial z}
\def\ppx#1{\partial #1/\partial x}
\def\ppy#1{\partial #1/\partial y}
\def\ppw#1{\partial #1/\partial w}
\def\ppr#1{\partial #1/\partial r}
\def\pprh#1{\partial #1/\partial \rho}
\def\pps#1{\partial #1/\partial s}
\def\ppt#1{\partial #1/\partial t}
\def\ppV#1{\partial #1/\partial V}
\def\pph#1{\partial #1/\partial \theta}
\def\ppvh#1{\partial #1/\partial \vartheta}
\def\ppxi#1{\partial #1/\partial \xi}

\def\ddt#1{d #1/dt}
\def\pppz#1{\partial^2 #1/\partial z^2}
\def\pppx#1{\partial^2 #1/\partial x^2}
\def\pppy#1{\partial^2 #1/\partial y^2}
\def\pppr#1{\partial^2 #1/\partial r^2}
\def\ppprh#1{\partial^2 #1/\partial \rho^2}
\def\ppps#1{\partial^2 #1/\partial s^2}
\def\pppt#1{\partial^2 #1/\partial t^2}
\def\pppV#1{\partial^2 #1/\partial V^2}
\def\ppph#1{\partial^2 #1/\partial \theta^2}
\def\pppvh#1{\partial^2 #1/\partial \vartheta^2}
\def\pppxi#1{\partial^2 #1/\partial \xi^2}
\def\dddt#1{d^2 #1/dt^2}

\def\grad{\nabla}
\def\cross{{\bf \times}}
\def\div{\grad\cdot}
\def\divp{\grad_\perp\cdot}
\def\divpl{\grad_\parallel\cdot}
\def\curl{\grad\cross}
\def\dpl{\grad_\parallel}
\def\ddpl{\grad_\parallel^2}
\def\dpp{\grad_\perp}
\def\ddpp{\grad_\perp^2}
\def\delsq{\grad^2}
\def\delamb{ \mathchar"0274\hskip -.665em\mathchar"0275 }
\let\delam=\delamb
\def\lapl{\grad^2}
\def\lapldef{\ddpp=(\pt^2/\pt x^2)+K^2(\pt^2/\pt y^2)}

\def\pwww#1{{\partial #1\over\partial \vec w}}
\def\pwwpl#1{{\partial #1\over\partial w_\parallel}}
 
\def\pvv#1#2{{\partial #2\over\partial v_{#1}}}
\def\ppv#1#2{{\partial #2/\partial v_{#1}}}
\def\pvvv#1{{\partial #1\over\partial \vec v}}
\def\pvvp#1#2{{\partial #2\over\partial v'_{#1}}}
\def\ppvp#1#2{{\partial #2/\partial v'_{#1}}}
\def\pvvvp#1{{\partial #1\over\partial \vec v'}}
\def\pvvpl#1{{\partial #1\over\partial v_\parallel}}
 
\def\xunit{\vec{\hat x}}
\def\yunit{\vec{\hat y}}
\def\zunit{\vec{\hat z}}
\def\sunit{\vec{\hat s}}
\def\bunit{\vec{b}}
\def\eunit{\vec{\hat e}}
\def\nunit{\vec{\hat n}}
\def\dt{\Delta t}
\def\becomes{\leftarrow}
\def\from{\leftarrow}
\def\to{\rightarrow}
\def\fromto{\leftrightarrow}
\def\implies{\,\,\,\Longrightarrow\,\,\,}
\def\dotdot{\!:\!}

\def\meters{\,{\rm m}}
\def\invm{\,{\rm m}^{-3}}
\def\invmeter{\,{\rm m}^{-1}}
\def\invsec{\,{\rm sec}^{-1}}
\def\cm{\,{\rm cm}}
\def\km{\,{\rm km}}
\def\invcc{\,{\rm cm}^{-3}}
\def\invcm{\,{\rm cm}^{-1}}
\def\invmm{\,{\rm mm}^{-1}}
\def\mm{\,{\rm mm}}
\def\Vcm{\,{\rm V/cm}}
\def\Acm{\,{\rm A/cm^2}}
\def\kA{\,{\rm kA}}
\def\MA{\,{\rm MA}}
\def\degk{\,{\rm K}}
\def\ergs{\,{\rm erg}}
\def\eV{\,{\rm eV}}
\def\keV{\,{\rm keV}}
\def\MeV{\,{\rm MeV}}
\def\GeV{\,{\rm GeV}}
\def\kG{\,{\rm kG}}
\def\tesla{\,{\rm T}}
\def\kW{\,{\rm kW}}
\def\MW{\,{\rm MW}}
\def\MWsqm{\,{\rm MW/m^2}}
\def\Wsqm{\,{\rm W/m^2}}
\def\radsec{\,{\rm rad/sec}}
\def\Hz{\,{\rm Hz}}
\def\kHz{\,{\rm kHz}}
\def\MHz{\,{\rm MHz}}
\def\mpersec{\,{\rm m}/{\rm sec}}
\def\msqsec{\,{\rm m^2}/{\rm sec}}
\def\cmsec{\,{\rm cm}/{\rm sec}}
\def\kmsec{\,{\rm km}/{\rm sec}}
\def\mmsec{\,{\rm m}^2/{\rm sec}}
\def\msqsec{\,{\rm m}^2/{\rm sec}}
\def\cmcmsec{\,{\rm cm}^2/{\rm sec}}
\def\ccpersec{\,{\rm cm}^3/{\rm sec}}
\def\minutes{\,{\rm min}}
\def\yr{\,{\rm yr}}
\def\hr{\,{\rm hr}}
\def\Bar{\,{\rm bar}}
\def\sec{\,{\rm sec}}
\def\msec{\,{\rm msec}}
\def\usec{\,\mu{\rm sec}}

\def\ee{\vec E}
\def\bb{\vec B}
\def\ff{\vec F}
\def\jj{\vec J}
\def\qq{\vec q}
\def\aa{\vec A}
\def\kk{\vec k}
\def\vv{\vec v}
\def\uu{\vec u}
\def\xx{\vec x}
\def\ww{\vec w}

\def\bdel{\vec b\cdot\grad}
\def\Bdel{\vec B\cdot\grad}
\def\Jdel{\vec J\cdot\grad}
\def\bdot{\vec B\cdot}
\def\Bdot{\vec B\cdot}
\def\kdot{\vec k\cdot}
\def\exb{\vec E\cross\vec B}
\def\jxb{\vec J\cross\vec B}
\def\uxb{\vec u\cross\vec B}
\def\vxb{\vec v\cross\vec B}
\def\wxb{\vec w\cross\vec B}
\def\ucxb{{\vec u\over c}\cross\vec B}
\def\vcxb{{\vec v\over c}\cross\vec B}
\def\wcxb{{\vec w\over c}\cross\vec B}
\def\jcxb{{\vec J\cross\vec B\over c}}

\def\vexb{\vec v_E}
\def\vpol{\vec v_p}
\def\upol{\vec u_p}
\def\vstar{\vec v_*}
\def\ustar{\vec u_*}
\def\Jstar{\vec J_*}
\def\Jpol{\vec J_p}
\def\vgradb{\vec v_{\grad B}}
\def\qpol{\vec q_p}
\def\qstar{\vec q_\wedge}
\def\qestar{\vec q_e{}_\wedge}
\def\qistar{\vec q_i{}_\wedge}
\def\pistar{\vec\Pi_*}
\def\vR{\vec v_R}
\def\vdl{\vec v\cdot\grad}
\def\vdel{\vec v\cdot\grad}
\def\vedl{\vexb\cdot\grad}
\def\udl{\vec u\cdot\grad}
\def\udel{\vec u\cdot\grad}
\def\uidl{\vec u_i\cdot\grad}
\def\uidel{\vec u_i\cdot\grad}
\def\wdel{\vec w\cdot\grad}
\def\dedt#1{d_E #1/dt}
\def\dett#1{{d_E #1\over dt}}
\def\jpp{J_\perp}
\def\jperp{\vec\jpp}
\def\qpp{q_\perp}
\def\qperp{\vec\qpp}
\def\upp{u_\perp}
\def\uperp{\vec\upp}
\def\wpl{w_\parallel}
\def\wpp{w_\perp}
\def\wperp{\vec\wpp}
\def\vpp{v_\perp}
\def\vperp{\vec\vpp}
\def\lnb{\log B}
 
\def\rms{_{rms}}
 
\def\Jpl{J_\parallel}
\def\jpl{J_\parallel}
\def\Jpp{J_\perp}
\def\jpp{J_\perp}
\def\Jperp{\vec\Jpp}
\def\Bperp{\vec B_\perp}
\def\Apl{A_\parallel}
\def\apl{A_\parallel}
\def\App{A_\perp}
\def\app{A_\perp}
\def\Aperp{\vec\App}
\def\Epl{E_\parallel}
\def\epl{E_\parallel}
\def\Epp{E_\perp}
\def\epp{E_\perp}
\def\Eperp{\vec\Epp}
\def\upl{u_\parallel}
\def\vpl{v_\parallel}
\def\Upl{U_\parallel}
\def\vor{\grad_\perp^2\phi}
\def\kpl{k_\parallel}
\def\kkpl{k_\parallel^2}
\def\kpp{k_\perp}
\def\kperp{\vec\kpp}
\def\kkpp{k_\perp^2}
\def\xpl{{x_\parallel}}
\def\xpp{x_\perp}
\def\DD{\Delta_D}
\def\Dpl{D_\parallel}
\def\Dpp{\Delta_\perp}
\def\Depl{D_e{}_\parallel}
\def\Dipl{D_i{}_\parallel}
\def\Rpl{R_\parallel}
\def\qpl{q_\parallel}
\def\qepl{q_e{}_\parallel}
\def\qipl{q_i{}_\parallel}
\def\Pipl{\Pi_\parallel}
\def\qeperp{\vec q_e{}_\perp}
\def\qiperp{\vec q_i{}_\perp}
\def\mupl{\mu_\parallel}
\def\mupp{\mu_\perp}
\def\nuei{\nu_{ei}}
\def\nuee{\nu_{ee}}
\def\nuii{\nu_{ii}}
\def\wpe{\omega_{pe}}
\def\wpi{\omega_{pi}}
\def\nudamp{\nu_d}
\def\zeff{Z_{\!e\!f\!f}}
\def\lmfp{\lambda_{\!m\!f\!p}}
\def\ws{{\omega_*}}
\def\wsi{{\omega_{*i}}}
\def\wn{\omega_n}
\def\wt{\omega_t}
\def\wi{\omega_i}
\def\wT{\omega_T}
\def\wp{\omega_p}
\def\wc{{\omega_c}}
\def\kappacv{{\cal K}}
\def\wcv{{\omega_B}}
\def\etai{\eta_i}
\def\taui{\tau_i}
\def\rs{\rho_s}
\def\ld{\lambda_D}
\def\Lpl{L_\parallel}
\def\Lpp{L_\perp}
\def\lcorpl{\lambda_\parallel}
\def\lcorpp{\lambda_\perp}
\def\lcorx{\lambda_x}
\def\lcory{\lambda_y}
\def\rch{\rho_{ch}}
\def\npl{\eta_\parallel}
\def\etapl{\eta_\parallel}
\def\ald{a_L}
\def\alde{a_{Le}}
\def\aldi{a_{Li}}
\def\npp{\eta_\perp}
\def\etapp{\eta_\perp}
\def\kappapl{\kappa_\parallel}
\def\dprime{\Delta'}
\def\sk{{}_{\vec k}}
\def\sky{{}_{k_y}}
\def\gk{\gamma_k}
\def\vk{\vfl_k}
\def\nk{\nfl_k}
\def\tk{\tfl_k}
\def\dk{\Delta k}
\def\gd{\gamma_0}
\def\mwn{\Delta_n}
\def\mwh{\Delta_h}
\def\gamT{\Gamma_T}
\def\gamn{\Gamma_n}
\def\gamt{\Gamma_t}
\def\gami{\Gamma_i}
\def\gamc{\Gamma_c}
\def\gamk{\Gamma_k}
\def\gams{\Gamma_s}
\def\gaml{\Gamma_l}
\def\gamr{\Gamma_r}
 
\def\ptb{\widetilde}
\def\psifl{\widetilde\psi}
\def\phifl{\widetilde\phi}
\def\ffl{\widetilde f}
\def\fe{f_e}
\def\fefl{\widetilde f_e}
\def\fifl{\widetilde f_i}
\def\nfl{\widetilde n}
\def\hfl{\widetilde h}
\def\tfl{\widetilde T}
\def\nefl{\widetilde n_e}
\def\nifl{\widetilde n_i}
\def\tefl{\widetilde T_e}
\def\tifl{\widetilde T_i}
\def\pfl{\widetilde p}
\def\pefl{\widetilde p_e}
\def\pifl{\widetilde p_i}
\def\hefl{\widetilde h_e}
\def\vx{\widetilde v_x}
\def\vfl{\widetilde v}
\def\vefl{\widetilde \vexb}
\def\vxfl{\widetilde v_x}
\def\vyfl{\widetilde v_y}
\def\vrfl{\widetilde v_r}
\def\vppfl{\widetilde v_\perp}
\def\vflpp{\widetilde v_\perp}
\def\vplfl{\widetilde \vpl}
\def\Bfl{\widetilde \vec B}
\def\Bflpp{\widetilde B_\perp}
\def\Aplfl{\widetilde A_\parallel}
\def\Appfl{\widetilde A_\perp}
\def\Aperpfl{\widetilde {\vec A}_\perp}
\def\ufl{\widetilde u_\parallel}
\def\vorfl{\grad_\perp^2\phifl}
\def\jfl{\widetilde J_\parallel}
\def\qfl{\widetilde q_\parallel}
\def\qefl{\widetilde q_e{}_\parallel}
\def\qifl{\widetilde q_i{}_\parallel}
\def\jppfl{\widetilde J_\perp}
\def\jperpfl{\widetilde {\vec J}_\perp}
\def\Afl{\ptb A_\parallel}
\def\Jfl{\ptb J_\parallel}
\def\efl{\widetilde E_\parallel}
\def\Efl{\widetilde E_\parallel}
\def\Eppfl{\widetilde E_\perp}
\def\Eperpfl{\widetilde {\vec E}_\perp}
\def\etafl{\widetilde\eta}
\def\isatfl{\widetilde I_{{\rm sat}}}
\def\phiflfl{\widetilde\phi_{{\rm fl}}}
 
\def\teplfl{\widetilde T_e{}_\parallel}
\def\teppfl{\widetilde T_e{}_\perp}
\def\qeplfl{\widetilde q_e{}_\parallel}
\def\qeppfl{\widetilde q_e{}_\perp}
\def\tiplfl{\widetilde T_i{}_\parallel}
\def\tippfl{\widetilde T_i{}_\perp}
\def\qiplfl{\widetilde q_i{}_\parallel}
\def\qippfl{\widetilde q_i{}_\perp}

\def\tepl{ T_e{}_\parallel}
\def\tepp{ T_e{}_\perp}
\def\qepl{ q_e{}_\parallel}
\def\qepp{ q_e{}_\perp}
\def\tipl{ T_i{}_\parallel}
\def\tipp{ T_i{}_\perp}
\def\qipl{ q_i{}_\parallel}
\def\qipp{ q_i{}_\perp}

\def\peplfl{\widetilde p_e{}_\parallel}
\def\peppfl{\widetilde p_e{}_\perp}
\def\piplfl{\widetilde p_i{}_\parallel}
\def\pippfl{\widetilde p_i{}_\perp}

\def\pepl{ p_e{}_\parallel}
\def\pepp{ p_e{}_\perp}
\def\pipl{ p_i{}_\parallel}
\def\pipp{ p_i{}_\perp}


\def\phinn{ {e\phifl\over T} }
\def\nnn{ {\nfl\over n} }
\def\tnn{ {\tfl\over T} }
\def\unn{ {\ufl\over c_s} }
\def\vornn{ \rho_s^2\ddpp\phinn }
\def\jnn{ {\jfl\over ne} }
\def\qnn{ {\qfl\over nT} }
\def\psinn{ {\psifl\over B\rho_s} }

\def\ahat{\hat\alpha}
\def\ehat{\hat\eta}
\def\khat{\hat\kappa}
\def\shat{\hat s}
\def\bhat{\hat\beta}
\def\muhat{\hat\mu}
\def\epss{\hat\epsilon}
\def\bigpoint#1{
    \par\bigskip
    {\baselineskip=\normalbaselineskip
    \parindent=0 pt
    {\hfill\vbox{ #1  }\hfill}}
    \par\bigskip
    }
 
\def\jfm#1{{\it J. Fluid. Mech.} {\secfnt #1}}
\def\jgr#1{{\it J. Geophys. Res.} {\secfnt #1}}
\def\prl#1{{\it Phys. Rev. Lett.} {\secfnt #1}}
\def\physletta#1{{\it Phys. Lett. A} {\secfnt #1}}
\def\physlettb#1{{\it Phys. Lett. B} {\secfnt #1}}
\def\pf#1{{\it Phys. Fluids} {\secfnt #1}}
\def\pfa#1{{\it Phys. Fluids A} {\secfnt #1}}
\def\pfb#1{{\it Phys. Fluids B} {\secfnt #1}}
\def\physp#1{{\it Phys. Plasmas} {\secfnt #1}}
\def\nf#1{{\it Nucl. Fusion} {\secfnt #1}}
\def\njp#1{{\it New J. Phys.} {\secfnt #1}}
\def\cpp#1{{\it Contrib. Plasma Phys.} {\secfnt #1}}
\def\ppcf#1{{\it Plasma Phys. Contr. Fusion} {\secfnt #1}}
\def\plasphys#1{{\it Plasma Phys.} {\secfnt #1}}
\def\revpp#1{{\it Rev. Plasma Phys.} {\secfnt #1}}
\def\iaea#1#2{in {\it Plasma Physics and Controlled Nuclear Fusion
    Research #1}, Proceedings of the #2th International Conference}
\def\EPS#1#2#3{in {\it Proceedings of the
{#1}th European Conference on Controlled Fusion and Plasma Physics,
{#2}, {#3}} (European Physical Society, {#2}, {#3})}
\def\jcp#1{{\it J. Comput. Phys.} {\secfnt #1}}
\def\jetp#1{{\it Sov. Phys. JETP} {\secfnt #1}}
\def\sovjpp#1{{\it Sov. J. Plasma Phys.} {\secfnt #1}}
\def\jnm#1{{\it J. Nucl. Mat.} {\secfnt #1}}
\def\rsi#1{{\it Rev. Sci. Inst.} {\secfnt #1}}
\def\adv#1{{\it Adv. Phys.} {\secfnt #1}}
\def\apjl#1{{\it Astrophys. J. Lett.} {\secfnt #1}}
\def\apj#1{{\it Astrophys. J.} {\secfnt #1}}
\def\astrap#1{{\it Astron. Astrophys.} {\secfnt #1}}
\def\mnras#1{{\it MNRAS} {\secfnt #1}}
\def\vol#1{\ {\secfnt #1}}

\def\itemc{\hfb \null\hskip 20 pt $\circ$\ }
\def\itemcont{\\ \hskip 40 pt }

\def\qqquad{\qquad\qquad}

\def\sumsp{\sum_{\hbox{sp}}}
\def\dL{d\Lambda\,}
\def\dW{d{\cal W}\,}
\def\dV{d{\cal V}\,}

\def\scripte{{\cal E}}

\def\Bpl{B^*_\parallel}
\def\Astar{\vec A^*}
\def\bstar{\vec b^*}
\def\Bstar{\vec B^*}
\def\Bstara{B^*{}^a}
\def\Bppa{\Bpp^a}
\def\pvvpl#1{{\pt #1\over\pt\vpl}}
\def\pmumu#1{{\pt #1\over\pt\mu}}
\def\pTT#1{{\pt #1\over\pt T}}
\def\pbb#1{{\pt #1\over\pt b}}

\def\magnet{{\cal M}}
\def\bstardot{\bstar\cdot}
\def\bstardel{\bstardot\grad}
\def\Bstardot{\Bstar\cdot}
\def\Bstardel{\Bstardot\grad}
\def\fhat{\vec{\hat F}}
\def\fdot{\fhat\cdot}
\def\fdel{\fdot\grad}

\def\bdot{\bunit\cdot}
\def\bdel{\bdot\grad}
\def\vor{\Omega}
\def\vorfl{\ptb\Omega}
\def\lcorx{\lambda_x}
\def\lcory{\lambda_y}

\def\bb{\vec B}
\def\uu{\vec u}
\def\fdot{\vec{\hat F}\cdot}
\def\fdl{\vec{\hat F}\cdot\grad}
\def\fdrxy{\hat F^{xy}}
\def\Bpp{B_\perp}
\def\vex{v_E^x}
\def\vey{v_E^y}
\def\bbx{b^x}
\def\bby{b^y}
\def\uex{u^x}
\def\uey{u^y}
\def\wex{w^x}
\def\wey{w^y}

\def\phig{\phi_G}
\def\vorg{\vor_G}
\def\phige{\phi_e}
\def\vorge{\vor_e}
\def\phigfl{\phifl_G}
\def\vorgfl{\vorfl_G}
\def\psig{A_G}
\def\chig{\chi_G}
\def\psigfl{\ptb A_G}
\def\chigfl{\ptb\chi_G}
\def\ufl{u_\parallel}
\def\Jfl{J_\parallel}

\def\Tpl{T_\parallel}
\def\Tpp{T_\perp}
\def\Tepl{T_e{}_\parallel}
\def\Tepp{T_e{}_\perp}
\def\Tplfl{\widetilde T_\parallel}
\def\Tppfl{\widetilde T_\perp}
\def\uepl{u_e{}_\parallel}
\def\qeplpl{q_e{}_\parallel{}_\parallel}
\def\qepppl{q_e{}_\perp{}_\parallel}

\def\ufl{\widetilde u_\parallel}
\def\tplfl{\widetilde T_\parallel}
\def\tppfl{\widetilde T_\perp}
\def\qplfl{\widetilde q_\parallel}
\def\qppfl{\widetilde q_\perp}
\def\pplfl{\widetilde p_\parallel}
\def\pppfl{\widetilde p_\perp}

\def\qplpl{\widetilde q_\parallel{}_\parallel}
\def\qpppl{\widetilde q_\perp{}_\parallel}

\def\tauz{\tau_z}
\def\rmuz{\mu_z}
\def\azz{a_z}

\def\nzfl{n_z}
\def\uzfl{u_z{}_\parallel}
\def\tzplfl{T_z{}_\parallel}
\def\tzppfl{T_z{}_\perp}
\def\qzplfl{q_z{}_\parallel}
\def\qzppfl{q_z{}_\perp}
\def\pzplfl{p_z{}_\parallel}
\def\pzppfl{p_z{}_\perp}
\def\Gzpl{G_z{}_\parallel}
\def\Gzpp{G_z{}_\perp}
\def\Wzpl{W_z{}_\parallel}
\def\Wzpp{W_z{}_\perp}
\def\Qzpl{Q_z{}_\parallel}
\def\Qzpp{Q_z{}_\perp}

\def\nefl{n_e}
\def\uefl{u_e{}_\parallel}
\def\teplfl{T_e{}_\parallel}
\def\teppfl{T_e{}_\perp}
\def\qeplfl{q_e{}_\parallel}
\def\qeppfl{q_e{}_\perp}
\def\peplfl{p_e{}_\parallel}
\def\peppfl{p_e{}_\perp}
\def\Gepl{G_e{}_\parallel}
\def\Gepp{G_e{}_\perp}
\def\Wepl{W_e{}_\parallel}
\def\Wepp{W_e{}_\perp}
\def\Qepl{Q_e{}_\parallel}
\def\Qepp{Q_e{}_\perp}

\def\nifl{n_i}
\def\uifl{u_i{}_\parallel}
\def\tiplfl{T_i{}_\parallel}
\def\tippfl{T_i{}_\perp}
\def\qiplfl{q_i{}_\parallel}
\def\qippfl{q_i{}_\perp}
\def\piplfl{p_i{}_\parallel}
\def\pippfl{p_i{}_\perp}
\def\Gipl{G_i{}_\parallel}
\def\Gipp{G_i{}_\perp}
\def\Wipl{W_i{}_\parallel}
\def\Wipp{W_i{}_\perp}
\def\Qipl{Q_i{}_\parallel}
\def\Qipp{Q_i{}_\perp}

\def\Nzfl{\LB n_z+\nzfl\RB}
\def\Tzplfl{\LB T_z+\tzplfl\RB}
\def\Tzppfl{\LB T_z+\tzppfl\RB}
\def\Pzplfl{\LB p_z+\pzplfl\RB}
\def\Pzppfl{\LB p_z+\pzppfl\RB}

\def\kkpp{k_\perp^2}

\def\uexb{\vec u_E}
\def\wexb{\vec w_E}
\def\Wexb{\vec W_E}

\def\uedl{\uexb\cdot\grad}
\def\wedl{\wexb\cdot\grad}
\def\Wedl{\Wexb\cdot\grad}

\def\bperp{\vec b_\perp}
\def\bbdl{\Bperp\cdot\grad}
\def\pxxmu#1{{\pt #1\over\pt x^\mu}}
\def\pxxnu#1{{\pt #1\over\pt x^\nu}}
\def\pyyk#1{{\pt #1\over\pt y_k}}
\def\ppyyk#1{{\pt^2 #1\over\pt y_k^2}}
\def\chiv{\hat\chi}

\def\kkpp{k_\perp^2}
\def\dpls{\dpl^0}

\section{Introduction}

Gyrofluid equations in toroidal geometry were originally derived for
small-amplitude disturbances to treat linear instabilities
\cite{Beer96}.  Nonlinear terms were taken from an earlier version
derived in slab geometry \cite{Dorland93}.  Landau damping dissipation,
resulting from phase mixing into arbitrarily small velocity space
structures, was taken from a previous closure dissipation model
\cite{Hammett90}.  Electromagnetic versions were also derived
\cite{aps99,Snyder01}.
The original idea behind gyrofluid equations, derived
in two-dimensional slab geometry, was for the density of a fluid of
gyrocenters rather than particles, with a polarisation equation tying
the gyrocenter densities together in a statement of overall charge
neutrality \cite{Knorr88}.  Polarisation results from the part of the
space density arising from the gyrophase angle dependent part of the
distribution function, while the gyrocenter density reflects the
part independent of the gyrophase angle.  This was already a result of
gyrokinetic theory \cite{Lee83}, which was originally formulated as a
finite Larmor radius (FLR) correction to the drift kinetic equation
\cite{FriemanChen82}.  The gyrofluid model rests on the gyrokinetic
model, much in the same fashion that the more familiar fluid model rests
on the Vlasov (or Boltzmann) equation for particles
\cite{Grad,Braginskii}.

The gyrokinetic theory itself, however, underwent a serious advance
resulting from the use of Lie transformation techniques applied to the
drift kinetic Lagrangian \cite{Littlejohn83,Dubin83,Hahm88}.  
The Lagrangian was
to be transformed to gyrocenter coordinates and only then the
gyrokinetic equation was to be found in terms of the resulting
Euler-Lagrange equations describing the motion of individual
gyrocenters.  The gyrofluid model, on the other hand, continued to be
based upon the older gyrokinetic formulation, in which it was not always
transparent what one had to keep in order to maintain consistency at a
given ordering.  Indeed, the original toroidal model had to be modified
before it had a clearly consistent energy conservation theorem
\cite{eps03,GEM}.  The gyrokinetic theory was re-cast as a Lagrangian field
theory \cite{Sugama00,Brizard00}, and nonlinear nonlocal gyrofluid
equations have also been formed in a similar manner
\cite{Strintzi04,Strintzi05}.  

The delta-f form of the gyrokinetic
theory also has its energy theorem \cite{LeeTang88}, whose conserved
quantity as a functional quadratic in all the dependent variables is
better thought of as a free energy, similar to that of fluid equations
\cite{Hasmim78,Turner82,WakHas84,ssdw}.  This has been
shown to be related to entropy \cite{Krommes94}.  Free energy for fluid
equations has also been shown to be related to this entropy
\cite{Sugama01}.  

The original gyrofluid derivation for toroidal geometry led to some
inconsistencies which caused fluctuation free energy not to be
conserved.  These were repaired by a construction method which can be
thought of as somewhat arbitrary and/or artificial \cite{GEM}.  
The free energy was
determined by analogy to the fluid equations, leaving it open to the
argument that it might not be fundamental.  The key insight in the
meantime was the connection made between the models, showing the
gyrofluid moment variables can be cast in terms of an Hermite polynomial
representation of the gyrokinetic distribution function.
The polynomials are functions of the velocity space coordinates, and the
coefficients of these are the gyrofluid moment variables.  Given the
gyrokinetic delta-f free energy functional, the gyrofluid one can be
derived using this Hermite expansion.  This was shown in Ref.\
\cite{Sugama01}\ in terms of a simplified gyrofluid model without FLR
corrections.

Herein, we systematise this procedure for the six-moment
gyrofluid model of Refs.\ \cite{Dorland93,Beer96,aps99} which was
corrected for energy conservation in Ref.\ \cite{GEM}.  But before that
we need to show the relation to the only really first-principle model in
the hierarchy: the total-f gyrokinetic equation
obtained by Lie transforms \cite{Dubin83,Hahm88}.
We start with that one, 
simplifying it mildly to obtain a computationally
tractable form.
Then we make the delta-f approximations in such a way
as to keep energetic consistency intact.  The total-f energy obtainable
from the Noether theorem \cite{Sugama00,Brizard00} is replaced by the
delta-f free energy functional referred to above.  This
delta-f equation and its free energy theorem then launch the derivation
of the gyrofluid equations in local form (fully nonlinear equations, but
constant background parameters, dependent on drift ordering).
At the kinetic level, the gyrocenter part of
the charge density in 
the polarisation equation is the same in the
total-f and delta-f versions, and arises
from the interaction term in the total-f Lagrangian.
The FLR closure for
the gyrofluid model is applied only once, and it is here.  
Ultimately, the role of polarisation is
the same in all three models, and taking that into account allows
energetic consistency to be maintained.  Specifically, 
the same rules must be applied both to polarisation and to the moment
variable equations \cite{GEM}.
The
polarisation part of the charge density is the same in all models which
use linearised polarisation, including the delta-f and gyrofluid levels.
This obtains
the polarisation equation, which at all levels can be used to re-cast the
ExB kinetic energy in terms of gyrocenter charge potential energy.
The part of
the gyrofluid free energy due strictly to the moment variables
follows from insertion of the
Hermite polynomial representation into the delta-f form.
The rest of the derivation is a mere matter of 
consistent application of the moments
to the delta-f gyrokinetic equations.  This part of the method is
similar to that of Refs.\ \cite{Dorland93,Beer96}, but the results of
Ref.\ \cite{GEM} 
concerning the conservative part of the equations and the free energy
are recovered without further pitfall.
Dissipation is inserted manually as before.  

Previous versions of gyrofluid equations have applied the FLR closure
only to the electrostatic potential, leaving the electromagnetic
response along.  This is usually justified since the turbulence spectrum
only extends down in scale as far as the ion gyroradius.  Hence, FLR
electron effects are small by the mass ratio.  Moreover, ion dynamical
contributions to the Ohm's Law are also finite mass ratio corrections.
However, fluid-type equations such as these are also useful in the study
of collisionless reconnection, which includes sufficiently small-scale
phenomena that electron FLR effects may enter
\cite{Porcelli91,Schep94,Grasso99,recon}.  Herein, the FLR treatment is
extended to the magnetic potential at the same level of sophistication
as to the electrostatic potential.  The parallel velocity and
perp/parallel heat flux moments are mixed in the magnetic flutter
disturbance and induction physics in the same way as the density and
perpendicular temperature moments in the ExB advection and polarisation
physics.  This extension of the gyrofluid model represents the new
result of this work.

\section{The total-f and delta-f forms of the gyrokinetic model
\label{sec:totalf}}


The total-f Vlasov equation is
a Hamiltonian bracket equation for the total distribution function $f$,
a dependent variable over phase space coordinates $\{\vec x,z,w\}$ with
the space part $\vec x$ given by field aligned coordinates $\{x,y,s\}$.
Field aligning refers to a single nonvanishing contravariant component
of the background magnetic field, in this case $B^s$, which identifies
$s$ as the parallel coordinate.  The velocity space part is given
by the parallel velocity $z$ and magnetic moment $w$, with $f$
independent of the gyrophase angle $\theta$.  
As a starting point we use the
result of the Lie transform theory version of the gyrokinetic 
expansion, specifically a version of the forms given in 
Eqs.\ (16,17,22--24) of Ref.\
\cite{Hahm88}, mildly simplified for computational use.
The particle Lagrangian is given by
\begin{equation}
L_p = {e\over c}\Astar\cdot\dtt{\vec x} + w{mc\over e}\dtt{\theta}-H
\label{eqlagrangian}
\end{equation}
where 
\begin{equation}
\Astar=\vec A+mz{c\over e}\bunit
\end{equation}
where $\vec A$ and $\bunit$ and $B$ are
the potential, unit vector, and field
strength of the background
magnetic field.  The Hamiltonian $H$ is given by
\begin{equation}
H=m{z^2\over 2}+wB+eJ_0\phi - {e^2\over 2B}\pww{}[J_0(\phi^2)-(J_0\phi)^2]
\label{eqhamiltonian}
\end{equation}
where $\phi$ is the electrostatic potential, $J_0$ is the gyroaveraging
operator, and the contribution which is quadratic in $\phi$ is referred
to as the gyroscreening potential.  Formally, $J_0$ has the form in
wavenumber space of multiplication of Fourier coefficients by
the zeroth Bessel function
$J_0(\kpp\rho_L)$, where $\rho_L$ is the particle gyroradius given by 
$\rho_L=\vpp/\Omega$, with gyrofrequency
$\Omega=\abs{eB/mc}$, or in terms of the coordinates by
$\rho_L^2=2w B/(m\Omega^2)$.
The particle equations of motion are
found from the Euler-Lagrange equations resulting from $L_p$,
\begin{equation}
\Bpl\dtt{\vec x} = z\Bstar - {c\over eB}\vec F\cdot\grad H \qquad\qquad 
	m\Bpl\dtt{z} = -\Bstardel H
\end{equation}
Drift tensor notation is used, with definitions
\begin{equation}
\Bstar=\curl\Astar \qquad\qquad  \Bpl=\bdot\Bstar
\end{equation}
\begin{equation}
\vec F = \grad\vec A-(\grad\vec A)^T
\end{equation}
It follows that
\begin{equation}
\vec F = \eps\cdot\vec B \qquad \curl\bunit = -\div{\vec F\over B}
\qquad\qquad\Bstar = \vec B - mz\div{c\over e}{\vec F\over B}
\end{equation}
where $\eps$ is the rank-three Levi-Civita pseudotensor.

The polarisation equation is found by extending the
Lagrangian to obtain one for the entire particles/field system using the
methods of Refs.\ \cite{Sugama00,Brizard00},
\begin{equation}
L = \sumsp\int\dL f\LB
        {e\over c}\Astar\cdot\dtt{\vec x} + w{mc\over e}\dtt{\theta}-H\RB
\label{eqfieldlagrangian}
\end{equation}
where the sum is over particle species,
$f$ is the distribution function for particles of each species,
$L_p$ multiplies $f$ expressed in phase space coordinates,
$H$ is given in Eq.\ (\ref{eqhamiltonian}), 
and the phase space, velocity space, and configuration space
integration domains are given respectively by 
\begin{equation}
\int\dL = \int\dV\int\dW \qqquad \dV = d^3x \qqquad 
\dW = 2\pi m^{-1}\, dz\,dw\,\Bpl
\label{eqdwelement}
\end{equation}
hence identifying $\Bpl$ as the variable part of the velocity space 
volume element.

The Euler-Lagrange equations for $\phi$ are found by varying the
Lagrangian with respect to $\phi$, yielding an integral over $\dV$ of
$\delta\phi(\vec x)$ times a coefficient, which is required to vanish.
This produces 
\begin{equation}
\sumsp\int\dW \LB eJ_0f+J_0\LP \magnet J_0\phi\RP
	- \LP J_0\magnet\RP\phi\RB = 0
\label{eqpolarisation}
\end{equation}
where $\magnet$ is the polarisability given by
\begin{equation}
\magnet = -{e^2\over B}\pww{f}
\end{equation}

It is customary in most treatments to linearise the polarisation term
involving $\magnet$.  This corresponds to replacing $f$ by a
Maxwellian background $F^M$ in the term in the system Lagrangian which
is due to the gyroscreening potential.  Since the dependence of $F^M$
upon $w$ is proportional to $\exp(-wB/T)$, the polarisability is
replaced by
\begin{equation}
\magnet \qquad\to\qquad e^2 F^M/T
\end{equation}
The Hamiltonian is replaced by 
\begin{equation}
H \qquad\to\qquad m{z^2\over 2}+wB+eJ_0\phi
\label{eqtotalh}
\end{equation}
and the polarisation equation is replaced by
\begin{equation}
\sumsp\int\dW \LB eJ_0f + e^2\,J_0\LP {F^M\over T} J_0\phi\RP 
	- \LP J_0{F^M\over T}\RP\phi\RB = 0
\label{eqtfpol}
\end{equation}
where now only the source term depends on the variable $f$.  
The facts that $H$ is linear
in $\phi$ and polarisation is linear in $f$ lead to the fact that this
particular set of terms remains intact when the delta-f approximations
are taken.  Ultimately, mutual field/particle energy conservation
depends on the properties of these terms, and the gyrofluid energy
considerations descend directly from the total-f gyrokinetic ones.

Most treatments neglect the action of $J_0$ upon $F^M$, leaving the
familiar form with $(e^2/T)F^M(J_0^2-1)\phi$, whose velocity space
integration produces $(ne^2/T)(\Gamma_0-1)\phi$, where $\Gamma_0$ is an
operator whose form in wavenumber space is multiplication of Fourier
coefficients by $I_0(b)e^{-b}$ with argument $b=\kkpp\rho_i^2$ evaluated
with the thermal gyroradius $\rho_i$, in turn given by 
$\rho_i^2=MTc^2/e^2B^2$ for species $i$ \cite{Lee83}.
In recent total-f gyrokinetic computations, this operator is further
simplified by using $\div (nMc^2/B^2)\grad\phi$, its low-$\kpp$ limit
\cite{Virginie07,Idomura07,Garbet07}.  
In these forms $n$ and $T$ are background
constants or, at most, radial (across magnetic flux surfaces) profiles
for the species density and temperature.

When polarisation is linearised, the corresponding term in the
Lagrangian becomes a field energy term, with $f$ replaced by $F^M$.
Hence the second order field piece (gyroscreening potential) 
is moved from $H$ into the field
energy term in the Lagrangian.  Consistency is maintained only when
these steps are taken together: if polarisation is linear then the
gyroscreening potential must be absent in $H$, or vice versa
\cite{Sugama00}.

The equation for $f$ is now determined by the particle equations of
motion with the linearised $H$, with the actual form remaining intact.
It is given by
\begin{equation}
\Bpl\ptt{f} + \grad H\cdot{c\over e}{\vec F\over B}\cdot\grad f
	+{1\over m}\Bstardot\LP\pzz{H}\grad f-\pzz{f}\grad H\RP = 0
\label{eqtotalf}
\end{equation}
Eqs.\ (\ref{eqtotalh},\ref{eqtfpol},\ref{eqtotalf}) form the
electrostatic version of
the total-f gyrokinetic model which is our starting point for the rest
of the derivation.  First, the simplest extension to include shear
Alfv\'en dynamics is given, then the delta-f model and its energy
theorem are presented, and then the gyrofluid model descends from that.

\subsection{Electromagnetic extension}

The discussion of the electromagnetic version of the model is kept brief
because the point is not to discuss various representations but to
establish what is needed to proceed to the delta-f formulation.  This
version retains $z$ as the parallel velocity coordinate, not the
canonical parallel momentum (cf.\ Ref. \cite{Hahm88a}).  The particle
Lagrangian $L_p$ retains the form in Eq.\ (\ref{eqlagrangian}) with
$\Astar$ now expanded to include the parallel magnetic potential,
$\Apl$, only, so that
\begin{equation}
\Astar=\vec A+\LP J_0\Apl+mz{c\over e}\RP\bunit
\end{equation}
The main complications are that the part multiplying $\bunit$ is
not only spatially variable but contains a time dependent field
variable.  The particle equations of motion become
\begin{equation}
\Bpl\dtt{\vec x} = z\Bstar - {c\over eB}\vec F\cdot\grad H \qquad\qquad 
	m\Bpl\dtt{z} = -\Bstardel H - {e\over c}\Bpl\,J_0\ptt{\Apl}
\end{equation}
with $\Bstar$ now containing magnetic nonlinearities involving
$\grad(J_0\Apl)$.  The overall Lagrangian including the fields becomes
\begin{equation}
L = \sumsp\int\dL f\LB
        {e\over c}\Astar\cdot\dtt{\vec x} + w{mc\over e}\dtt{\theta}-H\RB
        - \int\dV {1\over 8\pi}\abs{\dpp\Apl}^2
\end{equation}
with the last term representing the magnetic energy, and with $\Apl$
also occurring in $\Astar$.  Variation of this with respect to $\phi$ is
unchanged, and produces Eq.\ (\ref{eqpolarisation}) or its
simplifications such as Eq.\ (\ref{eqtfpol}).  Variation with respect to
$\Apl$ produces the induction equation (gyrokinetic version of Ampere's
law), 
\begin{equation}
\ddpp\Apl = -{4\pi\over c}\sumsp\int\dW \LB eJ_0 zf\RB
\label{eqinduction}
\end{equation}
where the term on the right gives the current, and without using the
canonical parallel momentum as a coordinate there is no complication
with 
the skin depth.  The latter does however remain implicitly present as
the electromagnetic gyrokinetic Vlasov equation for this Lagrangian,
\begin{equation}
\Bpl\ptt{f} - {e\over mc}\Bpl\pzz{f}J_0\ptt{\Apl}
        + \grad H\cdot{c\over e}{\vec F\over B}\cdot\grad f
        +{1\over m}\Bstardot\LP\pzz{H}\grad f-\pzz{f}\grad H\RP = 0
\label{eqtotalfem}
\end{equation}
now includes the time dependent induction effect through the explicit
time derivative of $\Apl$.

\subsection{The delta-f approximations
\label{sec:deltaf}}

The delta-f Vlasov equation results
from the expansion $f=F^M+\delta f$, with $\delta f$ a correction to a
Maxwellian background $F^M$ whose sole spatial dependence is through the
magnetic field strength $B$, and application of the delta-f ordering.
Prior to the application of strict delta-f ordering, the parallel
dynamics is linearised, so that the only nonlinear terms are those
resulting from $\grad H\cdot\vec F\cdot\grad$ or $\Bstardel$,
arising from the perturbed
contributions of $H$, $\Apl$ and $f$.  
With $\rho_i/L$ ordered small (with $L$
any background scale), $\Bpl$ is replaced by $B$.  Further, the
curvature is set such that $\bdel\bunit=\dpp\log B$.  This yields
\begin{eqnarray}
& &B\ptt{g} + \grad\psi_e\cdot{c\over B}\vec F\cdot\grad(\delta f)
	+ {mz^2+wB\over e}\grad\log B\cdot{c\over B}\vec F\cdot\grad h
	\nonumber\\
& &\qqquad{}+ {1\over m}\pzz{H_0}\Bdel{h} - {1\over m}\pzz{h}\Bdel{H_0} = 0
\label{eqgkdeltaf}
\end{eqnarray}
where the auxiliary variables are
\begin{equation}
  g = \delta f + {F^M\over T}e{z\over c}J_0\Apl \qquad\qquad
  h = \delta f + {F^M\over T}eJ_0\phi \qquad\qquad
  \psi_e = J_0\LP\phi-{z\over c}\Apl\RP
\label{eqdefgh}
\end{equation}
These are the inductive response, nonadiabatic response, and 
gyrokinetic potential,
respectively. 
The zeroth order Hamiltonian and the Maxwellian are given by
\begin{equation}
H_0 = m{z^2\over 2}+wB \qqquad
	F^M = n(2\pi T/m)^{-3/2}\exp(-H_0/T)
\label{eqdfmaxwellian}
\end{equation}
with $m$ and $e$ and $n$ and $T$ the species background parameters.

The terms in Eq.\ (\ref{eqgkdeltaf})
are referred to as nonlinear advection, magnetic drift, 
and parallel trapping and streaming, respectively.  
Nonlinear advection is by the perturbed Hamiltonian, which includes ExB
advection and the magnetic flutter nonlinearities, through $\phi$ and
$\Apl$, respectively.
The grad-B and curvature drifts are combined, and departures from an
inverse major radius dependence of $B$ are neglected.
The parallel dynamics includes both streaming ($\ppz{H_0}$) and magnetic
trapping ($\grad H_0 = wB\grad\log B$).  The parallel dynamics is
linearised, with both static and inductive pieces of $\Epl$ appearing
with $(mz/T)F^M$, leading to the expressions $h$ and $g$, respectively.

Now the strict form of delta-f ordering is applied.  This refers to the
split between the perpendicular and parallel coordinate directions
following from the $\kpl\ll\kpp$ ordering.  A field aligned coordinate
system is assumed, with $x$ and $y$ the perpendicular coordinates
respectively following the radial and electron drift directions, and 
the coordinate $s$ following the parallel direction.  This is done for a
general tokamak geometry following Refs.\
\cite{Beergeom,fluxtube,shifted}.  Axisymmetry renders $\ppy{}=0$ for
the background magnetic field $\vec B$.  The unperturbed parallel
gradient is given by
\begin{equation}
\Bdel = B^s\pss{} \qqquad\hbox{hence}\qquad\pxx{}\sim\pyy{}\gg\pss{}
\end{equation}
Under this ordering we have $B=B(s)$, and hence $B$ and $F^M$ are
independent of $x$ and $y$.  However, $\log B$ in the magnetic drift
terms must be re-formed in order to keep both the interchange
($\ppx{\log B}$) and geodesic ($\pps{\log B}$) curvature contributions. 
The derivatives of $\log B$ are evaluated, and then the resulting form
is restricted to have dependence upon $s$ only as with the rest of the
geometry.  
This introduces the curvature operator $\kappacv$, which is expressed
in terms of a vector-gradient contraction,
\begin{equation}
\kappacv \equiv \kappacv^x(s)\pxx{}+\kappacv^y(s)\pyy{}
\end{equation}
where the components are given by
\begin{equation}
\kappacv^{\{x,y\}} = -\grad\log B^2\cdot{c\vec F\over B^2}\cdot\grad\{x,y\}
\end{equation}
It is necessary for free energy conservation that $\kappacv$ be a pure
divergence.  This is guaranteed by maintaining
\begin{equation}
\pxx{}\kappacv^x + \pyy{}\kappacv^y = 0
\end{equation}
which is trivially satisfied by having the $\kappacv^a$ components
depend on $s$ only.  Correspondence to the linear forms in Refs.\
\cite{Dorland93,Beer96} is to identify $\kappacv$ with $-2i\omega_d$,
where $\omega_d$ is the toroidal drift frequency.

After this, the derivatives in the nonlinear advection term are
restricted to $\{x,y\}$ only, and the trapping/streaming terms involve
derivatives in $\{s,z\}$ only.  The resulting delta-f ordered
gyrokinetic equation is
\begin{equation}
\ptt{g}
	+ {c\over B_0}[(J_0\psi_e),h]_{xy}
	- {mz^2+wB\over 2e}\kappacv(h)
	+ {B^s\over mB}[H_0,h]_{zs} = 0
\label{eqdeltaf}
\end{equation}
where the brackets are given by
\begin{equation}
[f,g]_{ab} = {\pt f\over\pt x^a}{\pt g\over\pt x^b} 
	- {\pt f\over\pt x^b}{\pt g\over\pt x^a}
\end{equation}
with $a$ and $b$ denoting two of the phase space coordinates,
$\kappacv$ is the magnetic curvature operator.  The coefficients
$c$, $m$, $e$, and $B_0$ are constants; $x$, $y$, $s$, $z$, and $w$ are
coordinates; and $B$, the metric coefficients contained in $J_0$, and
the components of $\kappacv$ are functions of $s$ only.  The Maxwellian
$F^M$ is a function of $B$, $z$, and $w$ only.  Field aligned Hamada
coordinates constructed as in Refs.\ \cite{fluxtube,shifted}\ are
assumed.  Hence, $B^s$ and $F^{xy}/B^2$ are flux functions, which in the
local limit reduce to constants.  Following this is the use of the 
constant $B_0$, the average value of $B$ on a flux surface.

The field variables $\phi$ and $\Apl$ are functions of $\vec x$ and are
determined in the total-f theory by the polarisation and induction
equations, respectively.  In the total-f theory 
these equations are found by variation of the total action with respect
to the field variables.  The delta-f forms are the corresponding
linearised versions.  The polarisation equation is 
\begin{equation}
\sumsp\int\dW\LB e J_0 g + e^2{F^M\over T}(J_0^2-1)\phi\RB=0
\label{eqdfpol}
\end{equation}
resulting from Eq.\ (\ref{eqtfpol}).
The induction equation is
\begin{equation}
\ddpp\Apl + \sumsp{4\pi\over c}\int\dW\LB e z J_0 g 
	- {e^2\over c}z^2{F^M\over T}J_0^2\Apl\RB=0
\label{eqdfind}
\end{equation}
resulting from Eq.\ (\ref{eqinduction}).
In each case the delta-f form of the velocity space integral $\int\dW$
has reduced to
\begin{equation}
\dW = 2\pi m^{-1}\, dz\,dw\,B
\label{eqdfelement}
\end{equation}
The source terms in these equations follow directly from the lowest
order interaction Lagrangian terms,
\begin{equation}
\int\dL (\delta f)(-eJ_0\phi)  \qquad\qquad  
	\int\dL (\delta f){z\over c}(eJ_0\Apl)
\label{eqintlagrangian}
\end{equation}
respectively, where we can replace $f$ by $\delta f$ because these terms
are linear in $f$. 
The Hermitian property of $J_0$ and its commutation
with $z$ plays the central role.
This is why the way $J_0$ mixes
moments through its $w$-dependence must be established the same way for
the field variable derivatives in the Vlasov equation and the moment
source terms in the polarisation and induction equations.
The interchangeable
use of $g$ or $\delta f$ in Eq.\ (\ref{eqdfpol}) follows from the
antisymmetry of $zJ_0$ in velocity space.

\subsection{Energy and free energy in the gyrokinetic models
\label{sec:energy}}

The conserved energy $\scripte$ for the total-f model
is given by application of the Noether theorem to
its Lagrangian \cite{Sugama00,Brizard00}.  With $H$ and $L$ given in
Eqs.\ (\ref{eqhamiltonian},\ref{eqfieldlagrangian}), the energy is
merely the phase space integral of $H$ over $f$,
\begin{equation}
\scripte = \sumsp\int\dL Hf 
        = \sumsp\int\dL\LBR H_0 + eJ_0\phi 
        - {e^2\over 2B}\pww{}[J_0(\phi^2)-(J_0\phi)^2]\RBR f
\label{eqenergy1}
\end{equation}
The polarisation equation (Eq.\ \ref{eqpolarisation})
may be used to recast this in terms of
\begin{equation}
\scripte = \sumsp\int\dL\LBR H_0 f
        + {\magnet\over 2}[J_0(\phi^2)-(J_0\phi)^2]\RBR
\end{equation}
where the combination involving $\phi$ is the generalised ExB kinetic
energy (in the low-$\kpp$ limit it reduces to the familiar form
in terms of $v_E^2$, i.e., the usual ExB velocity squared).  Note the
use of the Hermitian property of $J_0$ as well as integration by parts
of $\ppw{}$.

In the form with linearised polarisation, the energy is equivalently
given by Eq.\ (\ref{eqenergy1}) with $f\to F^M$ in the gyroscreening
term.  Again the polarisation equation (here, Eq.\ \ref{eqtfpol}) is
used to eliminate the $feJ_0\phi$ term, to find
\begin{equation}
\scripte = \sumsp\int\dL\LB H_0 f 
        + e^2{F^M\over T}\LP 1-J_0^2\RP{\phi^2\over 2}\RB
\end{equation}
with $\magnet = (e^2/T)F^M$ inserted explicitly and the action of $J_0$
upon $F^M/T$ neglected.  The notation $J_0^2\phi^2$ is shorthand for
$(J_0\phi)^2$.  The velocity space integral over $F^M$ can be done
explicitly to find
\begin{equation}
\scripte = \sumsp\int\dL\LB H_0 f 
        + n{e^2\over T}\LP 1-\Gamma_0\RP{\phi^2\over 2}\RB
\end{equation}
where again $\Gamma_0\phi^2$ is shorthand for $\phi(\Gamma_0\phi)$.

This field energy term is the same in the delta-f model.  The
interaction Lagrangian term (implicit in the delta-f model) is also the
same, due to the linearity in both $f$ and $\phi$.  This
fact was used to motivate the discussion of free energy in delta-f
models generally \cite{LeeTang88,Krommes94,Sugama01}.  Here, however, we
have an even easier path, noticing that every term in Eq.\
(\ref{eqdeltaf}) is a bracket with $h$, i.e., a first order derivative
on $h$ and also a total divergence.  It is trivial to multiply Eq.\
(\ref{eqdeltaf}) by $h$ and any other quantity which commutes with all
the brackets to obtain conserved quantities, called Casimirs of the
system. 
The one identified with
free energy is quadratic in $h$ and must have dimensions of $nT$
integrated over space.  Hence the factor $(T/F^M)$ which works since
$F^M$ depends on $\{szw\}$ only and commutes with the $zs$-bracket, and
$T$ is constant.  The contribution quadratic in $\delta f$ is much like
a small-amplitude form of the thermodynamic entropy, while the
contributions due to $\phi$ and $\Apl$ become equivalent to the ExB and
magnetic field energies present in the total-f model, once the
polarisation and induction equations are used.
Details and underlying considerations are in Ref.\ \cite{Krommes94}.

The energy theorem for these equations
(\ref{eqdeltaf},\ref{eqdfpol},\ref{eqdfind}) is given by
\begin{equation}
\ptt{\scripte} 
	= \ptt{}\sumsp\int\dL\LB{T\over F^M}{hg\over 2}\RB
\label{eqdfenergya}
\end{equation}
or equivalently,
\begin{equation}
\ptt{\scripte} 
	= \ptt{}\sumsp\int\dL\LB{T\over F^M}{(\delta f)^2\over 2}
	+ e^2{F^M\over T}(1-J_0^2){\phi^2\over 2}\RB
	+ \ptt{}\int\dV{1\over 8\pi}\abs{\dpp\Apl}^2
\label{eqdfenergy}
\end{equation}
reflecting mutual conservation of thermal and kinetic free energy,
the E-cross-B energy, and the magnetic fluctuation energy, each given by
\begin{equation}
\scripte_f = \sumsp\int\dL {T\over F^M}{(\delta f)^2\over 2}
\label{eqdfenergyf}
\end{equation}
\begin{equation}
\scripte_E = \sumsp\int\dL e^2{F^M\over T}(1-J_0^2){\phi^2\over 2}
\label{eqdfenergye}
\end{equation}
\begin{equation}
\scripte_M = \int\dV {1\over 8\pi}\abs{\dpp\Apl}^2
\label{eqdfenergym}
\end{equation}
respectively.

The need to use the field equations to bring the factor of $h$ under the
$\ppt{}$ in Eq.\ (\ref{eqdfenergya}) is the reason the choice of
second-order Casimir to identify as the total free energy is unique.
Any integrable function of $h$ multiplying Eq.\ (\ref{eqdeltaf}) will
produce a vanishing phase space integral.  However, the operation
through Eqs.\ (\ref{eqdfpol},\ref{eqdfind}) will result in a single time
derivative only if the multiplier is $(T/F^M)h$.  This yields the
identification of $(T/F^M)hg/2$ as the free energy density and therefore 
$\scripte$ given by Eq.\ (\ref{eqdfenergy}) as the total free energy.

\section{The gyrofluid moment set and free energy}

The basic prescription of a gyrofluid model is the set of moments kept
as dependent variables.  The simplest version uses the density only
\cite{Knorr88}.  One level up from that is a three dimensional version
using densities and parallel velocities, equivalent in scope to the
familiar four field fluid models \cite{eps03}.  With the temperature
gradient of either species setting the basic dynamical character, at
least four moments, one each for the density $\nfl$, parallel velocity
$\ufl$, and parallel and perpendicular temperature $\Tplfl$ and $\Tppfl$, 
are needed \cite{Dorland93}.
These temperatures are given by the parallel and perpendicular energy
moments (over $mz^2/2$ and $wB$, respectively) divided by the density.
The perp/parallel separation in the temperatures is made necessary by
the underlying dynamics: only $\Tplfl$ enters parallel (hence Alfv\'en)
dynamics, and only $\Tppfl$ is involved in polarisation (through the
$wB$ dependence of $J_0$).  In the nonlinear dynamics
there is moment mixing between $\nfl$ and
$\Tppfl$ but not with $\Tplfl$, so even in the absence of curvature and
grad-B drifts the responses of $\Tppfl$ and $\Tplfl$ to the rest of the
dynamics are different.  Finally, magnetised plasma turbulence takes
place at time scales for which the parallel sound wave transit frequency
$c_s/qR$ is very slow.  Both the parallel viscosity and the
perp/parallel components of the parallel heat flux have time-dependent
responses to velocity and temperature gradients.  In a gyrofluid model
the parallel viscosity is proportional to the difference $\Tplfl-\Tppfl$,
which is already taken care of as $\Tplfl$ and $\Tppfl$ have their own time
dependent equations.  This means that the perp/parallel components of
the parallel heat flux ($\qplpl$ and $\qpppl$, moments over $mz^3/2$ and
$zwB$, respectively) 
also require their own time dependent equations.  The six moment models
\cite{Dorland93,Beer96,aps99,GEM} are the result.  

Here and below, the tilde symbols are used to distinguish the dependent
variables from the parameters.  There is one set of moment variables per
species, with all species contributing to polarisation and induction and
the field energy pieces.

The moment variable list is given by
\begin{equation}
\vbox{\hsize=6 cm
$$\nfl = \int\dW (\delta f)$$
$$n\ufl = \int\dW z\,(\delta f)$$
$$n\Tplfl = \int\dW (mz^2-T)\,(\delta f)$$
}\hskip 1 true cm
\vbox{\hsize=6 cm
$$n\Tppfl = \int\dW (wB - T)\,(\delta f)$$
$$\qplfl = \int\dW (mz^2-3T){z\over 2}\,(\delta f)$$
$$\qppfl = \int\dW (wB -T) z\,(\delta f)$$
}
\label{eqmomentsf}
\end{equation}
where here and below $\qplfl$ and $\qppfl$ are used as shorthand for
$\qplpl$ and $\qpppl$, respectively.  Heat fluxes perpendicular to $\vec
B$ are given by the curvature and grad-B drifts, as combinations of
$\nfl$, $\Tplfl$, and $\Tppfl$, and therefore are not written explicitly.

Use of a finite set of moment variables to represent $\delta f$ implies
a representation of $\delta f$ in terms of a finite-degree polynomial in
velocity space with those same moment variables as coefficients
\cite{Sugama01}.
The Hermite polynomial decomposition of the distribution
function in terms of the set of moment variables in Eqs.\ (\ref{eqmomentsf})
is given by
\begin{eqnarray}
\delta f = F^M\LB{\nfl\over n} + {\ufl\over V}\,{z\over V}
	+ \half{\Tplfl\over T}\LP{z^2\over V^2}-1\RP
	+ {\Tppfl\over T}\LP{wB\over mV^2}-1\RP
\right.\hskip 2 cm\nonumber\\\left.{}
	+ \third{\qplfl\over nTV}\LP{z^2\over V^2}-3\RP{z\over V}
	+ {\qppfl\over nTV}\LP{wB\over mV^2}-1\RP{z\over V}
\RB
\label{eqhermitef}
\end{eqnarray}
where $V$ is the species thermal velocity given by $V^2=T/m$.  The
coefficients are chosen for orthogonality and to recover the above
definitions of the moment variables.  Insertion of this form into the
portion of the delta-f free energy in Eq.\ (\ref{eqdfenergy}) dependent
on $\delta f$ yields the gyrofluid free energy.  The part dependent on
the state variables, $\nfl$ or $\Tplfl$ or $\Tppfl$, is the thermal free
energy, and the part dependent on
the flux variables, $\ufl$ or $\qplfl$ or $\qppfl$, is the kinetic free
energy.  The spatial density of the thermal free energy is
\begin{equation}
U_t = {nT\over 2}\LB\LP{\nfl\over n}\RP^2
	+\half\LP{\Tplfl\over T}\RP^2+\LP{\Tppfl\over T}\RP^2\RB
\label{eqthermalenergy}
\end{equation}
and the density of the kinetic free energy is 
\begin{equation}
U_v = {nT\over 2}\LB\LP{\ufl\over V}\RP^2
	+\twothirds\LP{\qplfl\over nTV}\RP^2+\LP{\qppfl\over nTV}\RP^2\RB
\label{eqkineticenergy}
\end{equation}
Together, $U_t+U_v$ represent the delta-f thermal free energy given by
$\scripte_f$ in Eq.\ (\ref{eqdfenergyf}), and are derived directly from
it. 

The field energy pieces are the same as in the delta-f gyrokinetic
version.  Evaluating the integral over $(1-J_0^2)F^M$ in Eq.\
(\ref{eqdfenergye}), the gyrofluid E-cross-B energy density is found.
The magnetic energy density carries over directly from Eq.\
(\ref{eqdfenergym}).  These are given by
\begin{equation}
U_E = \sumsp \LB ne^2{(1-\Gamma_0)\over T}{\phi^2\over 2}\RB
\qquad\qquad
U_M = {1\over 8\pi}\abs{\dpp\Apl}^2
\label{eqfieldenergy}
\end{equation}
respectively.  The operator $\Gamma_0$ reflects gyroscreening.  
As noted above, its form in 
wavenumber space is the function
\begin{equation}
\Gamma_0(b)\equiv I_0(b)e^{-b}
\qquad\qquad\hbox{with}\quad b=\kkpp\rho^2
\end{equation}
where $\rho=V/\Omega$ is the thermal gyroradius.  
Recall that in the local model $\Gamma_0$
involves derivatives with respect to $x$ and $y$ only, while $B=B(s)$ so
that the Hermitian property is maintained.  If wavenumber space is
unavailable, the Pad\'e approximant $\Gamma_0(b)=(1+b)^{-1}$ is used.
Eqs.\ (\ref{eqthermalenergy}-\ref{eqfieldenergy}) are the same as in
Ref.\ \cite{GEM}, but now they have a firm derivation in terms of the
delta-f gyrokinetic
version, applying the procedure of Ref.\ \cite{Sugama01} to the
result in Eq.\ (\ref{eqdfenergy}).

\section{Closure rules for gyroaveraging}

The evaluation of $J_0^2$ in polarisation was trivial because it is
integrated over $F^M$ and the integral of $J_0(b)^2 e^{-b}$ is well
known.  However, in the source term in polarisation (Eq.\ \ref{eqdfpol})
only one factor of $J_0$ appears.  A closure approximation for $\int\dW
J_0F^M$ is needed.  The one used previously was decided from the
properties of resulting linear eigenfunctions \cite{Dorland93,Beer96}.
However it is possible to evaluate this directly.  Taking the density
moment over $h$, we find
\begin{equation}
\int\dW h = \nfl + \int\dW {F^M\over T}eJ_0\phi 
\end{equation}
In the second term, $\phi$ is dependent on space only, and the fact that
under the delta-f model $F^M$ commutes with perpendicular spatial
derivatives may be used to do the velocity space integral separately.
This moment defines the basic gyroaveraging operator,
\begin{equation}
\Gamma_1\equiv {1\over n}\int\dW F^M J_0
\label{eqgamma1}
\end{equation}
which acts the same way
on any spatially dependent moment or field variable.  It is
merely a special function, as we find by inserting the form of $F^M$,
doing the integral over $z$, and defining $x=wB/T$,
\begin{equation}
\Gamma_1(b) = \int_0^\infty dx\,e^{-x} J_0(\sqrt{2bx})
\end{equation}
This form was used by Knorr \etal\ \cite{Knorr88}.
Conforming to the practice of Beer \etal\ \cite{Beer96},
however, we keep to the definition
\begin{equation}
\Gamma_1(b) \to \Gamma_0^{1/2}(b)
\end{equation}
and hence the operation of $\Gamma_1$ on a spatial variable is like that
of $\Gamma_0$ in wavenumber space, with
$\Gamma_0^{1/2}(b)$ is used rather than $\Gamma_0(b)$.
If wavenumber space is
unavailable, the Pad\'e approximant $\Gamma_0(b)=(1+b/2)^{-1}$ is used.
The label
$\Gamma_1$ is used for generality: any Hermitian operator is admissible
if consistency is the only requirement, so that if $F^M$ is
non-Maxwellian then some form other than $\Gamma_0^{1/2}$ is chosen.
These considerations are explained and justified in Ref.\ \cite{Beer96}.

In the
polarisation equation the source term is $\int\dW eJ_0(\delta f)$, which
contains moments over both unity and $wB$ times $F^M$.  The
corresponding closure approximation for this is
found by inserting Eqs.\ (\ref{eqhermitef}) into the factor of
$eJ_0(\delta f)$ in Eq.\ (\ref{eqdfpol}) to obtain
\begin{equation}
\int\dW eJ_0 (\delta f) = ne\LP\Gamma_1 {\nfl\over n}
	+ \Gamma_2{\Tppfl\over T}\RP
\end{equation}
where $\Gamma_2$ is given by
\begin{equation}
\Gamma_2 = \int\dW{wB-T\over T}F^M J_0 = T\pTT{}\int\dW F^M J_0
\label{eqgamma2}
\end{equation}
following both appearances of $T$ in the factors in $F^M$ which remain
after the integral over $z$ is done (cf.\ Eq.\ \ref{eqdfmaxwellian}).
It follows that whichever approximation is taken for $\Gamma_1$ we
always have 
\begin{equation}
\Gamma_2 = T\pTT{\Gamma_1} \to b\pbb{\Gamma_1}
\end{equation}
where the latter form is the one to use in wavenumber
space.

We refer to $\Gamma_1$ and $\Gamma_2$ as the first and second
gyroaveraging operators, or alternatively as the gyroaverging operator
and its first FLR correction, respectively.  For a six-moment model,
this is as far as the moment hierarchy goes.  If the moment over
$(wB)^2$ is a dependent variable in an extended version of the theory,
then there is a further third gyroaveraging operator involving second
derivatives of $\Gamma_1$ with respect to $b$, and so forth.  
Strictly speaking, the $(wB)^2$ moment will result in a form determined
by $\pt^2\Gamma_1/\pt(\log b)^2$.
However, in this case we have to replace this 4th moment with a form
mandated by energy conservation. 

Energy conservation in
the delta-f Vlasov equation works with $h$
appearing under all the derivatives and free energy evolution
determined by multiplication of the equation by $h$ and integrating over
phase space with sum over species 
(Eqs.\ \ref{eqdeltaf},\ref{eqdfenergya},\ref{eqdfenergy}).  
The corresponding form of this is
that the moments of $h$ appear under the derivatives in the gyrofluid
moment equations.  These moments are merely those over $\delta f$
together with contributions due to $J_0\phi$.  Only two of these
contributions are
nonzero, as the others vanish due to odd symmetry or orthogonality in
the integrals over $z$.  The surviving ones are
\begin{equation}
\int\dW h = \nfl + n{e\over T}\Gamma_1\phi
\qqquad
\int\dW{wB-T\over T}\,h = n{\Tppfl\over T} + n{e\over T}\Gamma_2\phi
\label{eqhfcombinations}
\end{equation}
Hence in the linear terms in the moment equations, derivatives of 
$\nfl+n(e/T)\Gamma_1\phi$ must appear in that combination, and the same
holds for $\Tppfl + e\Gamma_2\phi$.  This was the procedure by which
Ref.\ \cite{GEM} repaired gyrofluid energy conservation.
The higher moments (4th and 5th) over $h$
which occur in the equations are then
determined by the same ones over $\delta f$, which are straightforward.
Hence for the $(wB)^2$ moment we have
\begin{equation}
\int\dW(wB)^2\,h = 
	\int\dW(wB)^2\,(\delta f) 
		+ {e\over T}\LP\int\dW wB\,wB\,F^M J_0\RP\phi
\end{equation}
The first piece is found by straightforward evaluation
\begin{equation}
\int\dW(wB)^2\,(\delta f) = 2pT\LP{\nfl\over n}+2{\Tppfl\over T}\RP
\end{equation}
Hence this moment over $h$ must have the form found by combining
$\Gamma_1\phi$ with $\nfl$ and $\Gamma_2\phi$ with $\Tppfl$,
\begin{equation}
\int\dW(wB)^2\,h = 2pT\LB\LP{\nfl\over n}+{e\over T}\Gamma_1\phi\RP
	+2\LP{\Tppfl\over T}+{e\over T}\Gamma_2\phi\RP\RB
\end{equation}
Subtracting these two we find the required form,
\begin{equation}
{e\over T}\LP\int\dW wB\,wB\,F^M J_0\RP\phi
	= 2pT\,{e\over T}\LP\Gamma_1+2\Gamma_2\RP\phi
\end{equation}
and hence the requirement on the operators
\begin{equation}
{1\over n}\int\dW \LP{wB\over T}\RP^2 F^M J_0 
	= 2\LP\Gamma_1+2\Gamma_2\RP
\end{equation}
and also
\begin{equation}
{1\over n}\int\dW \LP{wB-T\over T}\RP^2 F^M J_0 
	= \LP\Gamma_1+2\Gamma_2\RP
\label{eqgamma3}
\end{equation}
This replaces the form which would be found from two derivatives applied
to $\Gamma_1$.  But that third form is only viable in the case the 4th
moments of $\delta f$ are retained as dynamical variables.  In the
six-moment model however the last equation above replaces that, and the
moment hierarchy is closed.  The closure of higher moments over $\delta
f$ is given by its form in Eq.\ (\ref{eqhermitef}), and then the higher
moments over $J_0$ are given by the requirement that the combinations in
Eq.\ (\ref{eqhfcombinations}) always appear intact in the moments over
$h$. 

Equivalently, we list the two approximations involved in the gyrofluid
model: (1) the choice of moments to keep as dynamical variables, and (2)
the form taken for $\Gamma_1$.  The rest of the model then follows by
simple evaluation constrained by energy conservation (keeping the pieces
of moments over $h$ together).

\section{Gyrofluid polarisation and relation to energy conservation}

We now return to the gyrofluid polarisation equation.  Insertion of
$\delta f$ from Eq.\ (\ref{eqhermitef}) into Eq.\ (\ref{eqdfpol}) and
using the definitions of $\Gamma_1$ and $\Gamma_2$ in Eqs.\
(\ref{eqgamma1},\ref{eqgamma2}), we have
\begin{equation}
\sumsp\LB ne\LP\Gamma_1 {\nfl\over n} + \Gamma_2{\Tppfl\over T}\RP
	+ ne^2{\Gamma_0-1\over T}\phi\RB = 0
\label{eqpol}
\end{equation}
As in the delta-f gyrokinetic polarisation equation (Eq.\
\ref{eqdfpol}), this is a statement of strict quasineutrality, with each
species charge density given by the gyrocenter part (moment variables)
and the polarisation part (due to the electrostatic potential).

In the total-f or delta-f gyrokinetic models, we can recover the
interaction Lagrangian in Eq.\ (\ref{eqintlagrangian}) by multiplying
the polarisation equation by $\phi$ and integrating over space
(essentially un-doing the steps by which the polarisation equation is
derived in the first place).  Under linearised polarisation the
interaction Lagrangian is the same in both the total- and delta-f
models, since the term is linear in both $f$ and $\phi$.  We can do the
same in the gyrofluid version, either operation by $\int\dV\phi\times$
or simply by inserting $\delta f$ from Eq.\ (\ref{eqhermitef}) into
Eq.\ (\ref{eqintlagrangian}).  In all these forms, the Hermicity of
$J_0$ and hence $\Gamma_1$ and $\Gamma_2$ has a central role.

For $\phi$ the result is
\begin{equation}
L_{{\rm int,}\phi} = -\sumsp\int\dV \LB
	e\phi\LP\Gamma_1\nfl+{n\over T}\Gamma_2\Tppfl\RP\RB
\end{equation}
whose variation with respect to $\phi$ recovers
the gyrocenter source terms in Eq.\ (\ref{eqpol}).
Applying the Hermitian property of the $\Gamma$'s, this is equivalent to
\begin{equation}
L_{{\rm int,}\phi} = -\sumsp\int\dV \LB
	e\LP\nfl \phig+{n\over T}\Tppfl\vorg\RP\RB
\end{equation}
where
\begin{equation}
\phig = \Gamma_1(\phi) = {1\over n}\int\dW F^M(J_0\phi) \qquad
	\vorg = \Gamma_2(\phi) = \int\dW {wB-T\over nT}F^M(J_0\phi)
\label{eqdefphig}
\end{equation}
are defined as the first and second gyroaveraged potentials,
respectively.
This is the underlying reason $\phig$ is associated with $\nfl$ and
$\vorg$ with $\Tppfl$ and, ultimately, why they must appear together in
derivatives representing energy transfer processes between the various
equations.  

The corresponding forms for $\Apl$ use the delta-f induction equation
(Eq.\ \ref{eqdfind}) and the part of Eq.\ (\ref{eqintlagrangian}) due to
$\Apl$, with the result
\begin{equation}
L_{{\rm int,}\Apl} = \sumsp\int\dV \LB
	n{e\over c}\Apl\LP\Gamma_1\ufl+\Gamma_2{\qppfl\over T}\RP\RB
\end{equation}
Variation of this with respect to $\Apl$ recovers the source term in the
gyrofluid Ampere's law,
\begin{equation}
\ddpp\Apl + {4\pi\over c}
\sumsp\LB ne\LP\Gamma_1\ufl + \Gamma_2{\qppfl\over T}\RP\RB = 0
\label{eqind}
\end{equation}
We can also find this by inserting the form for $\delta f$ in Eq.\
(\ref{eqhermitef}) into the gyrokinetic Ampere's law (Eq.\
\ref{eqdfind}) and evaluating the velocity space integrals. 
Using the Hermitian property of the $\Gamma$'s,
the electromagnetic interaction Lagrangian can also be re-cast as
\begin{equation}
L_{{\rm int,}\Apl} = \sumsp\int\dV \LB
	n{e\over c}\LP\ufl\psig+{\qppfl\over T}\chig\RP\RB
\end{equation}
where
\begin{equation}
\psig = \Gamma_1(\Apl) = {1\over n}\int\dW F^M(J_0\Apl) \qquad
	\chig = \Gamma_2(\Apl) = \int\dW {wB-T\over nT}F^M(J_0\Apl)
\label{eqdefpsig}
\end{equation}
are defined as the first and second gyroaveraged magnetic potentials,
respectively.
This leads to the appearance of $\ufl$ and $\psig$ together, and
$\qppfl$ and $\chig$, under $\ppt{}$ in the gyrofluid moment equations.

All of this is closely related to conservation of free energy in terms
of the functionals in Eqs.\ (\ref{eqdfenergya},\ref{eqdfenergy}) 
for the delta-f gyrokinetic model
and Eqs.\ (\ref{eqthermalenergy}--\ref{eqfieldenergy}) for the gyrofluid
case.  The field energy components in Eqs.\ (\ref{eqfieldenergy}) can be
re-cast using the polarisation and induction equations in Eqs.\
(\ref{eqpol},\ref{eqind}) as 
\begin{equation}
\scripte_E=\sumsp\half\int\dV\LB 
	e\LP\phig\nfl+n\vorg{\Tppfl\over T}\RP\RB
\end{equation}
and
\begin{equation}
\scripte_M=\sumsp\half\int\dV\LB 
	{e\over c}\LP n\psig\ufl+\chig{\qppfl\over T}\RP\RB
\end{equation}
respectively.  The time derivatives follow accordingly, and the time
derivative of the total energy is given by
\begin{eqnarray}
\ptt{\scripte}=\sumsp\int\dV\LB\LP ne\phig+T\nfl\RP\ptt{}{\nfl\over n}
	+ ne\ufl\ptt{}\LP{1\over c}\psig+{m\over e}\ufl\RP
	+ \half{n\over T}\Tplfl\ptt{\Tplfl}
\right.\nonumber\\  {}+\left.{}
	{n\over T}\LP e\vorg+\Tppfl\RP\ptt{\Tppfl}
	+ \twothirds{m\over nT}\qplfl\ptt{}{\qplfl\over nT}
	+ e{\qppfl\over T}\ptt{}
		\LP{1\over c}\chig+{m\over e}{\qppfl\over nT}\RP\RB
\end{eqnarray}
which is the direct correspondence to the $h(\ppt{g})$ form in Eq.\
(\ref{eqdfenergy}) for the delta-f model.  This is the same energy
equation as  given in Ref.\ \cite{GEM}, but now it is a result, not a
construction.   

Following these we can find the moments over $h$ and $g$ defined in
Eqs.\ (\ref{eqdefgh}) by inserting Eq.\ (\ref{eqhermitef}) for $\delta
f$ and then straightforward evaluation of the velocity
space integrals using the definitions in Eqs.\
(\ref{eqdefphig},\ref{eqdefpsig}) for the potentials.
The moment list for $h$ in entirety is
\begin{equation}
\vbox{\hsize=7 cm
$$\nfl + n{e\over T}\phig = \int\dW h$$
$$n\ufl = \int\dW z\,h$$
$$n\Tplfl = \int\dW (mz^2-T)\,h$$
\vfill}\hskip 0.1 true cm
\vbox{\hsize=7 cm
$$n\Tppfl + ne\vorg = \int\dW (wB - T)\,h$$
$$\qplfl = \int\dW (mz^2-3T){z\over 2}\,h$$
$$\qppfl = \int\dW (wB -T) z\,h$$
\vfill}
\label{eqmomentsh}
\end{equation}
similar to Eqs.\ (\ref{eqmomentsf}) as only $\nfl$ and $\Tppfl$ are
affected.  The moment list for $g$ in entirety is
\begin{equation}
\vbox{\hsize=7 cm
$$\nfl = \int\dW g$$
$$n\ufl + n{e\over mc}\psig = \int\dW z\,g$$
$$n\Tplfl = \int\dW (mz^2-T)\,g$$
\vfill}\hskip 0.1 true cm
\vbox{\hsize=7 cm
$$n\Tppfl = \int\dW (wB - T)\,g$$
$$\qplfl = \int\dW (mz^2-3T){z\over 2}\,g$$
$$\qppfl + nT{e\over mc}\chig = \int\dW (wB -T) z\,g$$
\vfill}
\label{eqmomentsg}
\end{equation}
where due to the extra factor of $z$ it is $\ufl$ and $\qppfl$ which are
affected.  The moments over $g$ are the quantities appearing under
$\ppt{}$ in the gyrofluid equations, while those over $h$ appear under
the derivatives in the linear terms.

The salient result of this section is the way the interaction Lagrangian
and the Hermitian property fix the gyroaveraging operations for the
potentials once they are decided for the moment variables, and vice
versa.  This relationship is what guarantees an energy conserving model
using this particular procedure.  And given these constraints, the free
energy theorem follows naturally.

\section{The gyrofluid moment equations}

With the above in place, the straightforward derivation of the gyrofluid
moment equations is essentially determined.  The form of $\delta f$ in
terms of the moment variables, the definitions of the latter (hence
which moments to take), and the closure rules are already defined.  We
simply take the moment list defined in Eqs.\ (\ref{eqmomentsf}) and
apply each one separately to Eq.\ (\ref{eqdeltaf}).  With one exception,
all the steps follow directly.  The exception is the parallel magnetic
nonlinearities, in which both field variables appear, each with a factor
of $J_0$, and the velocity space integral must be decided.  But as we
will see, the form this must have is already determined by the
requirement of energy conservation.

\subsection{Toroidal curvature, magnetic divergence effects}

The delta-f form of the gyrokinetic equation splits the curvature and
grad-B drifts from the rest, as these are purely linear terms.  These
two drifts are combined, such that in each case we take a combined
moment over $mz^2+wB$.  The quantity operated upon is $h$, not $\delta
f$, so that the field potential moments also appear.  Only $\phi$ is
involved, and $J_0$ appears only with $\phi$.  Moreover, only
derivatives over $x$ and $y$ are involved, while $B$ depends only on
$s$.  These terms are therefore found by simple evaluation of the
moments over $mz^2+wB$ times $h$.  These terms conserve energy
separately from those arising from other brackets, as in the delta-f
Vlasov equation (in Eq.\ \ref{eqdeltaf} $h$ combines with itself under
$\kappacv$ as a
pure divergence).  The 4th moment ($w^2B^2$) over $J_0\phi$ in Eq.\
(\ref{eqgamma3}) is determined by this requirement, as noted there.
After evaluation of the moments, the curvature terms appear as
\begin{eqnarray}
&& \ptt{}{\nfl\over n} = \cdots + {T\over e}\kappacv\LP
	{\pplfl+\pppfl\over 2 nT}+{e\phig\over T}+{e\vorg\over 2T}\RP\\
&& \ptt{}\LP{1\over c}\psig+{m\over e}\ufl\RP 
	= \cdots + {mT\over e^2}\kappacv\LP
		2\ufl+{2\qplfl+\qppfl\over 2nT}\RP\\
&& \half\ptt{}{\Tplfl\over T} = \cdots + {T\over e}\kappacv\LP
	{\pplfl\over 2 nT}+{e\phig\over 2T}+{\Tplfl\over T}\RP\\
&& \ptt{}{\Tppfl\over T} = \cdots + {T\over e}\kappacv\LP
	{\pppfl\over 2 nT}+{e\phig+e\vorg\over 2T}
		+3{\Tppfl+e\vorg\over 2T}\RP\\
&& \ptt{}\LP{m\over e}{\qplfl\over nT}\RP 
	= \cdots + {mT\over e^2}\kappacv\LP
		\threehalves\ufl+4{\qplfl\over nT}\RP\\
&& \ptt{}\LP{1\over c}\chig+{m\over e}{\qppfl\over nT}\RP 
	= \cdots + {mT\over e^2}\kappacv\LP
		\half\ufl+3{\qppfl\over nT}\RP
\end{eqnarray}
The terms with $\kappacv$ operating directly on the moment variables
always appear with factors of $T/e$ in these scaled units, reflecting
the charge separation effect of ``diamagnetic'' curvature terms.  Those
acting on $\phi$ give the ExB compression effects plus FLR corrections.
In the parallel flux variable equations there is the additional factor
of $m/e$ matching the one in the inertia terms.
These curvature terms form a closed set which conserves energy
separately, as detailed in Ref.\ \cite{GEM}.

\subsection{Nonlinear ExB advection}

In the nonlinear brackets $[(J_0\phi),(\delta f)]_{xy}$
both $\phi$ and $\delta f$ appear under gradients.  The
moment mixing for state variables is given by
\begin{equation}
\int\dW \LB(J_0\phi),(\delta f)\RB_{xy} = 
	\LB\phig,\nfl\RB_{xy} 
		+ {n\over T}\LB\vorg,\Tppfl\RB_{xy}
\end{equation}
\begin{equation}
\int\dW (wB-T)\LB(J_0\phi),(\delta f)\RB_{xy} = 
	n\LB\phig,\Tppfl\RB_{xy} 
		+ \LB\vorg,\LP T\nfl+2n\Tppfl\RP\RB_{xy}
\end{equation}
For flux variables it is similar,
\begin{equation}
\int\dW z\LB(J_0\phi),(\delta f)\RB_{xy} = 
	n\LB\phig,\ufl\RB_{xy} 
		+ \LB\vorg,{\qppfl\over T}\RB_{xy}
\end{equation}
\begin{equation}
\int\dW z(wB-T)\LB(J_0\phi),(\delta f)\RB_{xy} = 
	\LB\phig,\qppfl\RB_{xy} 
		+ \LB\vorg,\LP nT\ufl+2\qppfl\RP\RB_{xy}
\end{equation}
For the pure parallel velocity moments there is no mixing as no factors
of $wB$ appear,
\begin{equation}
\int\dW (mz^2-T)\LB(J_0\phi),(\delta f)\RB_{xy} 
	= n\LB\phig,\Tplfl\RB_{xy} 
\end{equation}
\begin{equation}
\int\dW {z\over 2}(mz^2-T)\LB(J_0\phi),(\delta f)\RB_{xy} 
	= \LB\phig,\qplfl\RB_{xy} 
\end{equation}
The nonlinearities involving $\Apl$ properly belong to the parallel
dynamics.  Since the parallel gradient ultimately acts on $h$ rather
than $\delta f$ in the kinetic model, these nonlinearities involve two
appearances of the field potentials ($\Apl$ and $\phi$, in the nonlinear
parallel electric field).  In the ExB advection terms these pieces
vanish because they all arise from $[(J_0\phi),(J_0\phi)]_{xy}$.
Indeed, if we apply the rules for $h$ to the above six combinations
($\phig$ with $\nfl$ and $\vorg$ with $\Tppfl$, as in Eq.\
\ref{eqhfcombinations}), all the terms quadratic in $\phi$ vanish, as
they should do.  But for $[(J_0\Apl),(J_0\phi)]_{xy}$ several field
terms survive (the lowest order among them being the magnetic
flutter effect on the electric field in the Ohm's law, in the fluid
sense) and the energy theorem's constraints are needed to evaluate them.
They are left to the discussion on nonlinear parallel dynamics, which
follows next.

\subsection{Parallel dynamics and magnetic nonlinearities}

In the gyrokinetic nonlinear bracket both $\phi$ and $\Apl$ appear in
the gyrokinetic potential (perturbed Hamiltonian).  
Since $\Apl$ appears with an extra factor of
$z$, we compute these terms separately as they involve different
moments.  The lowest order $\Apl$ terms combine in the nonlinear
parallel gradient,
\begin{equation}
\dpl = {B^s\over B}\pss{} - [\psig,]_{xy}
\end{equation}
to which the $\Gamma_2$-dependent FLR effects add and mix moments in the
same way as with $\phig$ and $\vorg$ in the ExB advection.
However, the $\pps{}$ terms also involve the
dependence $B=B(s)$, so it is useful to consider these separately.

\subsection{Linear parallel dynamics}

The linear terms arise from the $[H_0,h]_{zs}$ bracket in Eq.\
(\ref{eqdeltaf}), whose two pieces combine parallel streaming and
magnetic trapping effects.  Here, $B$ does not commute with $\pps{}$ so
there arise extra terms proportional to $\pps{B}$, referred to as
magnetic pumping terms, the vestige of kinetic trapping in the gyrofluid
model \cite{Beer96}.
There is also the distinction between a parallel divergence and
a parallel gradient, according to whether $1/B$ occurs inside or outside
of $\pps{}$.  As with the curvature terms, the moments are over $h$, not
$\delta f$, so the same combinations arise as in the curvature terms.
With these extra considerations the moment calculations are
straightforward.  Again, 
these terms conserve energy separately and hence can be considered
separately (with the magnetic pumping terms independent of the others).
After evaluation of the moments, the linear parallel
gradient/divergence and magnetic pumping terms appear as
\begin{eqnarray}
&& \ptt{}{\nfl\over n} = \cdots - B\dpls{\ufl\over B}\\
&& \ptt{}\LP{1\over c}\psig+{m\over e}\ufl\RP 
	= \cdots - \dpls{ne\phig+T\nfl+n\Tplfl\over ne}
		-{(e\vorg+\Tppfl)-\Tplfl\over e}\dpls\log B\\
&& \half\ptt{}{\Tplfl\over T} = \cdots 
	- B\dpls{nT\ufl+\qplfl\over nTB}
		-{nT\ufl+\qppfl\over nT}\dpls\log B\\
&& \ptt{}{\Tppfl\over T} = \cdots - B\dpls{\qppfl\over nTB}
		+{nT\ufl+\qppfl\over nT}\dpls\log B\\
&& \ptt{}\LP{m\over e}{\qplfl\over nT}\RP 
	= \cdots - \threehalves\dpls{\Tplfl\over e}\\
&& \ptt{}\LP{1\over c}\chig+{m\over e}{\qppfl\over nT}\RP 
	= \cdots - \dpls{e\vorg+\Tppfl\over e}
		-{(e\vorg+\Tppfl)-\Tplfl\over e}\dpls\log B
\end{eqnarray}
where $\dpls$ denotes the linear part of the parallel gradient.  In
field-aligned Hamada coordinates it is given by
\begin{equation}
B\dpls = B^s\pss{}
\end{equation}
with $B^s$ independent of $s$.  The Jacobian does not appear because it
is a flux function (function of $x$ only).  Under strict delta-f
ordering both the Jacobian and $B^s$ are constants.
In general the Jacobian enters, so that
\begin{equation}
\dpls = {B^s\over B}\pss{}  \qquad\qquad B\dpls{f\over B} 
	= {1\over\sqrt{g}}\pss{}\LP\sqrt{g}{B^s\over B}f\RP
\end{equation}
where $g$ is the determinant of the metric coefficients (all three
components).  It is essential that $\sqrt{g}B^s$ be a flux function, to
preserve $\div \vec B=0$.

\subsection{Nonlinear parallel dynamics}

The lowest-order
nonlinear terms follow by consistency from the linear ones,
as both pieces of the nonlinear parallel derivative 
$\dpl=\dpls-[\psig,]_{xy}$ act together.
However, treating temperature dynamics with FLR consistency adds FLR
nonlinearities to these, involving brackets with $\chig$.  These involve
moment mixing in the same way as for $\phig$ and $\vorg$ in the ExB
advection.  The moment integrals encountered are basically the same,
because the integrals over $z$ and $w$ separate, and the parallel
dynamics is merely one order higher by the factor of $z$ in the moment
hierarchy.

Terms arising from $[(J_0\Apl),(\delta f)]_{xy}$ involving $\psig$ 
simply follow from the linear
ones arising from $\dpls(\delta f)$.  Those involving $\chig$ raise the
moment level by one in $w$ in the same way as in the ExB advection
terms.  The only subtlety is the one involving moments over the field
nonlinearity bracket $[J_0\Apl,J_0\phi]_{xy}$ which is unique to the
nonlinear parallel dynamics (the corresponding terms in ExB advection
vanish trivially with $[J_0\phi,J_0\phi]_{xy}$).  These terms appear in the
flux variable equations,
\begin{eqnarray}
&& \ptt{}\LP{1\over c}\psig+{m\over e}\ufl\RP 
	= \cdots + \int\dW mz\, {F^M\over nT}\, 
	{z\over B_0}[J_0\Apl,J_0\phi]_{xy} \\
&& \ptt{}\LP{m\over e}{\qplfl\over nT}\RP 
	= \cdots + \int\dW{mz^2-T\over T}\,{mz\over 2}\, {F^M\over nT}\, 
	{z\over B_0}[J_0\Apl,J_0\phi]_{xy} \\
&& \ptt{}\LP{1\over c}\chig+{m\over e}{\qppfl\over nT}\RP 
	= \cdots + \int\dW{wB -T\over T}\, mz\, {F^M\over nT}\, 
	{z\over B_0}[J_0\Apl,J_0\phi]_{xy}
\end{eqnarray}
since in the state variable equations they vanish due to the
odd symmetry of $\int\dW z$.
The difficulty is that the velocity space integration is no longer
associated with only one of the quantities appearing under spatial
derivatives in the bracket.  The solution is to apply the moments to the
state variable terms in the $(\delta f)$ representation in Eq.\
(\ref{eqhermitef}) first.  These give
\begin{eqnarray}
&& \ptt{}\LP{1\over c}\psig+{m\over e}\ufl\RP 
	= \cdots + {1\over ne B_0}[\psig,\pplfl]_{xy} 
		+ {1\over e B_0}[\chig,\Tppfl]_{xy} \\
&& \ptt{}\LP{m\over e}{\qplfl\over nT}\RP 
	= \cdots + {3\over 2eB_0}[\psig,\Tplfl]_{xy} \\
&& \ptt{}\LP{1\over c}\chig+{m\over e}{\qppfl\over nT}\RP 
	= \cdots + {1\over eB_0}[\psig,\Tppfl]_{xy}
		+ {1\over ne B_0}[\chig,(\pplfl+2n\Tppfl)]_{xy}
\end{eqnarray}
Now we apply the combination rules for $h$ given in Eq.\
(\ref{eqhfcombinations}), so that these become
\begin{eqnarray}
&& \ptt{}\LP{1\over c}\psig+{m\over e}\ufl\RP 
	= \cdots + {1\over ne B_0}[\psig,(\pplfl+ne\phig)]_{xy} 
\nonumber\\ && \qquad\qquad{}
		+ {1\over e B_0}[\chig,(\Tppfl+e\vorg)]_{xy} \\
&& \ptt{}\LP{m\over e}{\qplfl\over nT}\RP 
	= \cdots + {3\over 2eB_0}[\psig,\Tplfl]_{xy} \\
&& \ptt{}\LP{1\over c}\chig+{m\over e}{\qppfl\over nT}\RP 
	= \cdots + {1\over eB_0}[\psig,(\Tppfl+e\vorg)]_{xy}
\nonumber\\ && \qquad\qquad{}
		+ {1\over ne B_0}[\chig,(\pplfl+ne\phig)]_{xy}
		+ {2\over e B_0}[\chig,(\Tppfl+e\vorg)]_{xy}
\end{eqnarray}
Hence we have determined that the rules for moments over two operations
by $J_0$ on field variables are
\begin{eqnarray}
&&{1\over n}\int\dW F^M[(J_0\Apl),(J_0\phi)] 
	= [(\Gamma_1\Apl),(\Gamma_1\phi)] 
	+ [(\Gamma_2\Apl),(\Gamma_2\phi)] \\
&&{1\over n}\int\dW {wB\over T}F^M[(J_0\Apl),(J_0\phi)] 
	= [(\Gamma_1+\Gamma_2)\Apl,\Gamma_2\phi)] 
	+ [\Gamma_2\Apl,(\Gamma_1+\Gamma_2)\phi]
\end{eqnarray}
It is simple to show that these vanish if $\Apl\to\phi$, as they should
do.

In the state variable equations the $\Apl,\phi$ combinations do not
appear, and the nonlinear parallel divergences of flux variables are
left as
\begin{eqnarray}
&& \ptt{}{\nfl\over n} 
	= \cdots + {1\over B_0}[\psig,\ufl]_{xy} 
		+ {1\over nT B_0}[\chig,\qppfl]_{xy} \\
&& \half\ptt{}{\Tplfl\over T} 
	= \cdots + {1\over nT B_0}[\psig,(nT\ufl+\qplfl)]_{xy} 
		+ {1\over nT B_0}[\chig,\qppfl]_{xy} \\
&& \ptt{}{\Tppfl\over T} 
	= \cdots + {1\over nT B_0}[\psig,\qppfl]_{xy} 
		+ {1\over nT B_0}[\chig,(nT\ufl+2\qppfl)]_{xy}
\end{eqnarray}
Here, there is crosstalk between $\nfl$ or $\Tplfl$ 
and $\qppfl$ due to the fact
that the corresponding $wB-T$ moment does not vanish when a factor of
$J_0$ is present.  Again, this set of 
terms conserves energy separately
from all the others.  By contrast to ExB advection, however, the
conservation is between state and flux variable sets, not just within
each of those sets.  
When the flux variables are kept as time dependent variables, the
level of moment mixing within state and flux variable sets remains
consistent (hence velocity with density and heat flux with temperature,
with one heat flux for each of perpendicular and parallel temperatures).
With these terms determined, the dissipation free part of the equations,
the part which conserves free energy exactly, is now closed.  Under the
strict delta-f ordering, with the geometry dependent upon $s$ and with
nonlinear derivatives only in $\{x,y\}$, the constraint $\div\vec B=0$
is maintained in the nonlinear terms as well.

\section{Dissipative Effects}

The addition of dissipation is essentially done by hand, as in most of
the moment approaches (cf.\ Ref.\ \cite{Beer96}), as it is not part of
the original Lagrangian/Hamiltonian formulation.  Even at the
gyrokinetic level, if collisional effects are very weak then either the
computation must resolve very thin striations in phase space (cf.\ Refs.\
\cite{Shoucri77,Shoucri78,Ghizzo,Bertrand90}), or it must cut them off
via a hyperdiffusion in velocity space which accounts for what is
essentially the same thing as a ``high Reynolds number'' situation
viz-a-viz Fokker-Planck collisional diffusion
\cite{Dannert04,Dannert05}.  Here we note that the basic mathematical
properties of the one-dimensional (1D) Vlasov-Maxwell system of the earlier
references are the same as the shear-Alfv\'en subset of the
electromagnetic gyrokinetic turbulence discussed in the more recent
ones.  The general gyrokinetic collisional process is being developed on
more firm mathematical grounds \cite{Brizard04,Mishchenko07}, but it is
mostly simplified models that are currently in use \cite{hmode05}, when
collisions are used at all.

Dissipation effects due to both Landau damping and collisions are added
to the gyrofluid equations at more or less the same level of
sophistication.  While it is possible to mimic the linear response of
the plasma dispersion function in the linear limit in homogeneous
geometry \cite{Hammett90,Dorland93} or for simplified instabilities in
toroidal geometry \cite{Beer96}, a general fitting approach was shown to
fail even for linear instabilities close to and away from threshold, by
the necessity to use a different fitting matrix for each case
\cite{Beer96}.  Hence, Landau damping was inserted into the heat flux
equations as a direct-damping model in Ref.\ \cite{GEM}, such that the
Alfv\'en damping response is adequately captured without impacting any
of the conservative transfer processes (i.e., by changing their
coefficients, as was done in Ref.\ \cite{Dorland93}).

Collisions are another matter, as there is a well formulated fluid limit
to which the equations should relax if the collisional frequency becomes
large compared to any parallel transit or nonlinear advection
frequencies --- the Braginskii equations \cite{Braginskii}.  Typical
fluid models in the drift frequency regime keep the parallel dissipation
effects (resistivity, thermal forces, parallel viscosity)
\cite{dalfloc,Xu97,Rogers98}.  Gyrofluid equations, functioning in the
same regime, keep these same processes as well \cite{aps99,GEM}.  The
only complication is to keep them consistent with the anisotropic
temperature model, and to formulate the thermal forces such that they
disappear naturally when the collision frequency drops to zero
\cite{dalfloc}.  At the drift kinetic 
or gyrokinetic level, the dissipation is covered by the 1-D part of the
equations describing parallel dynamics, and a simple procedure has been
shown for the drift kinetic equation for electrons in Ref.\
\cite{Hassam80} which we can use here.  This is to set up a
Chapman-Enskog expansion around a simple Lorentz collision operator,
obtain directly all the necessary terms, and then change the
coefficients such that the result agrees with the Braginskii model.
Here, the only extension of that was to start from a bi-Maxwellian with
given perpendicular and parallel temperatures
\begin{equation}
F^{(0)} = n(2\pi \Tpl/m)^{-1/2}(2\pi \Tpp/m)^{-1}\exp(-wB/\Tpp-mz^2/2\Tpl)
\end{equation}
and then to solve the correction equation
\begin{equation}
\LP z\dpl + {e\over m}\Epl\pzz{}\RP F^{(0)} 
	= {\nu_L\over v^3}\pzeze{}(1-\zeta^2)\pzeze{f^{(1)}}
\end{equation}
for $f^{(1)}$, where coordinates $\{v,\zeta\}$ are given by
$mv^2=2wB+mz^2$ and $\zeta=z/v$, and $\nu_L$ is the Lorentz collision
parameter. 
Following Ref.\ \cite{Hassam80}, at lowest order there are no flows (in
the local rest frame of the ion fluid), so the consistency conditions by
which $\ppt{}$ is eliminated are trivial (small parallel force
imbalance, no divergences).
Also, a finite $\Apl$ is neglected by assuming the resistivity is
sufficiently large, and toroidal drifts are neglected by assuming
$R\gg\Lpp$, where $\Lpp$ is the profile scale length.  Having solved for
$f^{(1)}$, the heat flux variables $\qplpl$ and $\qppfl$ and, for
electrons, the parallel current $\Jpl=-n_e e\uefl$ are
evaluated directly.  This yields collisional formulae for the flux
variables 
\begin{eqnarray}
& &\eta{m_e\nu_e\over e}\Jpl
	= n_e e\Epl+\dpl\pepl+\alpha_e n_e\dpl\Tepl \\
& &\qeplpl+1.28(\qeplpl-1.5\qepppl)
	+{3\over 5}\alpha_e {T_e\over e}\Jpl 
	= -{3\over 5}\kappa_e{T_e\over m_e\nu_e}n_e\dpl\Tepl\\
& &\qepppl-1.28(\qeplpl-1.5\qepppl)
	+{2\over 5}\alpha_e {T_e\over e}\Jpl 
	= -{2\over 5}\kappa_e{T_e\over m_e\nu_e}n_e\dpl\Tepp\\
\end{eqnarray}
as well as anisotropy dissipation corrections to the temperature
equations 
\begin{eqnarray}
\label{eqtiso}
& &\half n_e\ptt{\Tepl} + \cdots = -{\nu_e\over 3\pi_e}(\Tepl-\Tepp)\\
& & n_e\ptt{\Tepp} + \cdots = {\nu_e\over 3\pi_e}(\Tepl-\Tepp)
\label{eqtiso2}
\end{eqnarray}
Here, the numerical coefficients resulting from the Lorentz model which
are kept are written explicitly, while the coefficients to be
substituted with their Braginskii values are written as $\eta$,
$\alpha_e$, $\kappa_e$, and $\pi_e$, for resistivity, thermoelectric
coupling, thermal conduction, and (eventually) parallel viscosity,
respectively.  As the subscripts indicate, the calculation itself is
done for electrons.  For ions, the thermoelectric coupling is set to
zero, and then $\kappa_i$ and $\pi_i$ are given their Braginskii values
(for several ion species if desired).

To treat the transcollisional situation the thermoelectric coupling is
reformulated (obviously, it should vanish as $\nu_e\to 0$).  The
thermal force itself, $\alpha_e n_e\dpl\Tepl$, is substituted using the
heat flux formulae (in this, the anisotropy is assumed to be small for
large $\nu_e$), as detailed in Ref.\ \cite{dalfloc}, so that the
electron force imbalance is entirely due to dissipation of the flux
variables,
\begin{equation}
n_e e\LP{1\over c}\ptt{\Apl}+\dpl\phi\RP-\dpl\pepl 
	= -m_e\nu_e\LB\eta{\Jpl\over e}
	+{\alpha_e\over\kappa_e} 
	\LP{\qeplpl+\qepppl\over T_e}+\alpha_e{\Jpl\over e}\RP\RB
\label{eqjiso}
\end{equation}
having 
The heat flux formulae are treated in a similar fashion,
\begin{eqnarray}
\label{eqqiso}
& &\threehalves p_e\dpl\Tepl = -m_e\nu_e{5/2\over\kappa_e}
	\LB \qeplpl+1.28(\qeplpl-1.5\qepppl)
	+{3\over 5}\alpha_e {T_e\over e}\Jpl\RB\\
& &p_e\dpl\Tepp = -m_e\nu_e{5/2\over\kappa_e}
	\LB \qepppl-1.28(\qeplpl-1.5\qepppl)
	+{2\over 5}\alpha_e {T_e\over e}\Jpl\RB
\label{eqqiso2}
\end{eqnarray}
with gradients on the left and dissipative damping terms on the right.
The right hand sides of Eqs.\ (\ref{eqtiso}--\ref{eqqiso2})
represent a set of terms which are
added to the right hand sides to the corresponding
gyrofluid moment equations.  
The resistivity combination in Eq.\ (\ref{eqjiso})
subtracts from the right hand side of
$\ppt{\psig}$.
The temperature terms in Eqs.\
(\ref{eqtiso},\ref{eqtiso2}) add to the temperature equations as shown.
The heat flux terms in Eqs.\ (\ref{eqqiso},\ref{eqqiso2}) add to the
equations for $\qeplpl$ and $\qepppl$, respectively.  The ions are done
the same way as the electrons except for the resistivity terms (which
are the same for all species), and with the coefficients $\kappa_i$ and
$\pi_i$ the appropriate ones for each ion species, with $\alpha$ set to
zero for the ions.

Parallel viscosity does not explicitly appear in the equation for the
parallel velocity moment but instead results from collisional
dissipation of the difference between perpendicular and parallel
temperatures.  This difference is of course one and the same with the
parallel-parallel component of the viscous tensor \cite{Braginskii}.
Either one keeps a viscosity term in the parallel momentum equation or
one keeps track of thermal anisotropy with the dependent variables, but
not both.  There is some discussion of this in Refs.\
\cite{Dorland93,Beer96}, and it is the ultimate reason that no Landau
closure dissipative term should appear in the parallel velocity equation
itself \cite{Hammett90}.  Such a two moment dissipative closure has been
given \cite{Snyder01}, but especially for electrons it has the
undesirable property of mimicking a resistivity which is a factor of
$V_e/\nu_e qR$ too large.  Hence the Landau closure itself appears only
in the heat flux moment equations, following Refs.\
\cite{Dorland93,Beer96,GEM}, and the parallel viscosity is naturally
given by the collisional dissipation of thermal anisotropy if the
collision frequency is sufficiently dominant \cite{pet05}.  The
Chapman-Enskog procedure gives the coefficient $\nu_z/3$ for species
$z$, as in Ref.\ \cite{Beer96}.  The coefficient $\pi_z$ then gives the
correct viscosity coefficient for species $z$ according to the
collisional fluid derivation \cite{Braginskii}.  

The correspondence of these equations to low frequency fluid drift
equations including nonlinear polarisation and
collisional dissipation processes in the appropriate 
(``Braginskii'') regime has been shown elsewhere and a summary is given
in Sec.\ \ref{sec:braginskii}, below.

\par\vfill\eject

\section{Resulting gyrofluid equations}

The results of the above derivation are essentially the same as those of
Ref.\ \cite{GEM}.  The two new elements are the derivation path and the
inclusion of direct finite gyroradius effects in the part of the
nonlinear dynamics representing the fluctuations in the magnetic field.
The derivation is now firmly grounded within the underlying gyrokinetic
theory to the extent that the representation of the gyrokinetic
distribution function in terms of the gyrofluid moment variables is
explicit and the energy theorem not only remains intact but has itself
been used to determine the closure rules.  The part of the gyrofluid
moment equations which arises directly from the gyrokinetic model and is
exactly conservative is given by equations for the gyrocenter density,
\begin{equation}
\dtt{\nzfl} + [\vorg,\tzppfl]
	+ B\dpl{\uzfl\over B} - \beta_e[\chig,\qzppfl]
	= \kappacv\LP\tauz{\pzplfl+\pzppfl\over 2}+\phig+{\vorg\over 2}\RP
\label{eqnfl}
\end{equation}
parallel velocity,
\begin{eqnarray}
&& \beta_e\ptt{\psig}+\rmuz\dtt{\uzfl} + \rmuz[\vorg,\qzppfl]
	+ \dpl(\phig+\tauz\pzplfl) - \beta_e[\chig,(\vorg+\tauz\tzppfl)]
\nonumber\\ && \qquad{}
	+ (\vorg+\tauz\tzppfl-\tauz\tzplfl)\dpl\log B
	= \tauz\rmuz\kappacv\LP{4\uzfl+2\qzplfl+\qzppfl\over 2}\RP
\label{equfl}
\end{eqnarray}
parallel temperature,
\begin{eqnarray}
&& \half\dtt{\tzplfl} 
	+ B\dpl{\uzfl+\qzplfl\over B} - \beta_e[\chig,\qzppfl]
	+ (\uzfl+\qzppfl)\dpl\log B
\nonumber\\ && \qquad{}
	= \kappacv\LP\tauz{\pzplfl+2\tzplfl\over 2}+{\phig\over 2}\RP
\label{eqtplfl}
\end{eqnarray}
perpendicular temperature,
\begin{eqnarray}
&& \dtt{\tzppfl} + [\vorg,(\nzfl+2\tzppfl)]
\nonumber\\ && \qquad{}
	+ B\dpl{\qzppfl\over B} - \beta_e[\chig,(\uzfl+2\qzppfl)]
	- (\uzfl+\qzppfl)\dpl\log B
\nonumber\\ && \qquad{}
	= \kappacv\LP\tauz{\pzppfl+3\tzppfl\over 2}+{\phig+4\vorg\over 2}\RP
\label{eqtppfl}
\end{eqnarray}
parallel/parallel heat flux,
\begin{equation}
\rmuz\dtt{\qzplfl} 
	+ \threehalves\tauz\dpl\tzplfl
	= \tauz\rmuz\kappacv\LP{3\uzfl+8\qzplfl\over 2}\RP
\label{eqvqplfl}
\end{equation}
and perp/parallel heat flux,
\begin{eqnarray}
&& \beta_e\ptt{\chig}+\rmuz\dtt{\qzppfl} + \rmuz[\vorg,(\uzfl+2\qzppfl)]
\nonumber\\ && \qquad{}
	+ \tauz\dpl\tzppfl 
		- \beta_e[\chig,(\phig+\tauz\pzplfl)] 
		- \beta_e[\chig,2(\vorg+\tauz\tzppfl)]
\nonumber\\ && \qquad{}
	+ (\vorg+\tauz\tzppfl-\tauz\tzplfl)\dpl\log B
	= \tauz\rmuz\kappacv\LP{\uzfl+6\qzppfl\over 2}\RP
\label{eqvqppfl}
\end{eqnarray}
The lowest order nonlinear advective and parallel derivatives are
given by
\begin{equation}
\dtt{}=\ptt{}+[\phig,\,] \qqquad \dpl=\bunit\cdot\grad-\beta_e[\psig,\,]
\end{equation}
with $\bunit$ the unit vector of the unperturbed magnetic field.

The nonlinear brackets are given by
\begin{equation}
[f,g] = \grad f\cdot\vec F_0\cdot\grad g 
	\qqquad\hbox{given}\quad\div\vec F_0 = 0 
	\qquad\hbox{and}\quad\vec F_0\dotdot\grad\grad = 0 
\label{eqbracket}
\end{equation}
where $\vec F_0$ is a divergence free, antisymmetric tensor as
specified, in the particular normalisation being used.  Similarly for
the curvature operator,
\begin{equation}
\kappacv(f) = \kappacv^i\grad_i f
	\qqquad\hbox{given}\quad\grad_i\kappacv^i = 0 
\end{equation}
A typical case
is to leave the drift scale ratio $\rho_s/\Lpp$ out of the normalisation
and put it into $\vec F_0$ and $\kappacv$.  Conventional normalisation
is to fold it into the normalisation, keep to strict fluxtube ordering,
so that $[f,g]=f_{,x}g_{,y}-f_{,y}g_{,x}$ in the conventional linearised
gyro-Bohm version as in Refs.\ \cite{Dorland93,Beer96}.
Both versions are covered in Ref.\ \cite{GEM}.

The FLR reduced potentials are given by
\begin{equation}
\phig=\Gamma_1\phi \qqquad \psig=\Gamma_1\Apl
\label{eqphig}
\end{equation}
\begin{equation}
\vorg=\Gamma_2\phi \qqquad \chig=\Gamma_2\Apl
\label{eqvorg}
\end{equation}
in terms of the field potentials $\phi$ and $\Apl$.
The associated field potential equations are given by
\begin{equation}
\sum_z\azz\LB\Gamma_1\nzfl+\Gamma_2\tzppfl+{\Gamma_0-1\over\tauz}\phi\RB=0
\label{eqphi}
\end{equation}
for polarisation and
\begin{equation}
\ddpp\Apl + \sum_z\azz\LB\Gamma_1\uzfl+\Gamma_2\qzppfl\RB = 0
\label{eqpsi}
\end{equation}
for induction.  In a computational model the induction equation is
actually solved using the combinations under the $\ppt{}$ in 
Eqs.\ (\ref{equfl},\ref{eqvqppfl})
\begin{eqnarray}
&& 
\LP\sum_z\azz\LB {\beta_e\over\rmuz}\LP\Gamma_1^2+\Gamma_2^2\RP\RB - \ddpp\RP\Apl 
\nonumber\\ && \qqquad{}
	= \sum_z{\azz\over\rmuz}\LB\Gamma_1\LP\beta_e\psig+\rmuz\uzfl\RP
	+\Gamma_2\LP\beta_e\chig+\rmuz\qzppfl\RP\RB
\label{eqpsig}
\end{eqnarray}
as these quantities on the right hand side are what are actually
advanced.

The constant parameters
\begin{equation}
\azz = {n_zZ\over n_e} \qquad
\tauz = {T_z\over ZT_e} \qquad
\rmuz = {m_z\over Zm_D}
\end{equation}
given the background charge density, temperature/charge, and mass/charge
ratios, normalised to electron and deuterium values.  The pressures are
linearised, so that
\begin{equation}
\pzplfl = \nzfl+\tzplfl \qqquad \pzppfl = \nzfl+\tzppfl
\end{equation}
under gradient operators.  The profile gradients for all the state
variables are included as part of the dependent variables; these may be
split in a traditional manner without loss of generality.

To these equations are added the collisional dissipation model,
\begin{eqnarray}
&& \beta_e\ptt{\psig}+\rmuz\ptt{\uzfl} = \cdots
	- \mu_e\nu_e\LB\eta\Jpl+{\alpha_e\over\kappa_e}\LP
		\qeplfl+\qeppfl+\alpha_e\Jpl\RP\RB
\\
&& \half\ptt{\tzplfl} = \cdots - {\nu_z\over 3\pi_z}(\tzplfl-\tzppfl)
\\
&& \ptt{\tzppfl} = \cdots + {\nu_z\over 3\pi_z}(\tzplfl-\tzppfl)
\\
&& \rmuz\ptt{\qzplfl} = \cdots - \rmuz\nu_z{5/2\over\kappa_e}
	\LB \qzplfl+1.28(\qzplfl-1.5\qzppfl)
	+{3\over 5}\alpha_z\Jpl\RB
\\
&& \rmuz\ptt{\qzppfl} = \cdots - \rmuz\nu_z{5/2\over\kappa_e}
	\LB \qzppfl-1.28(\qzplfl-1.5\qzppfl)
	+{2\over 5}\alpha_z\Jpl\RB
\end{eqnarray}
with parallel current
\begin{equation}
\Jpl = \sumsp\azz\uzfl
\end{equation}
and with numerical coefficients
\begin{equation}
\alpha_e = 0.71 \qquad \kappa_e = 3.2
\qquad \pi_e = 0.73
\qquad \eta = 0.51
\end{equation}
for electrons and
\begin{equation}
\alpha_i = 0. \qquad \kappa_i = 3.9
\qquad \pi_i = 0.96
\end{equation}
for singly charged ions.  
Note the appearance of $\npl$ in the $\uzfl$-equation for all
species, since the resistivity essentially adds to $\ppt{\psig}$, and
that the thermoelectric coupling between $\Jpl$ and the heat fluxes
affects only the electrons.  For other charge states the corresponding
coefficients for ions may be found in Ref.\ \cite{Braginskii}.

Finally, the Landau damping model is added,
\begin{eqnarray}
&& \rmuz\ptt{\qzplfl} = \cdots 
	- \rmuz\sqrt{\tauz/\rmuz}\LP 1-0.125q^2R^2\ddpl\RP\qzplfl
\\
&& \rmuz\ptt{\qzppfl} = \cdots 
	- \rmuz\sqrt{\tauz/\rmuz}\LP 1-0.125q^2R^2\ddpl\RP\qzppfl
\end{eqnarray}
which gives a finite-difference compatible version of the original by
Hammett and Perkins \cite{Hammett90} and their successors
\cite{Dorland93,Beer96}, as explained in Ref.\ \cite{GEM}.  This
completes the description of the six-moment gyrofluid model
(``GEM'' from Refs.\ \cite{aps99,eps03,GEM}),
now extended to incorporate finite gyroradius effects in the nonlinear
magnetic fluctuation dynamics.

\subsection{Gyrofluid equations for collisionless reconnection}

Simplified two dimensional models are often used in studies of
collisionless reconnection
\cite{Porcelli91,Ottaviani93,Schep94,Ottaviani95,Grasso99,Fitzpatrick04,recon,Fitzpatrick07}. 
In terms of the geometry only the dynamics perpendicular to a prescribed
guide field is retained; in the language of this work this means only
the nonlinear brackets are kept, with the linear parallel derivative
incorporated into the magnetic nonlinearities, assuming a Cartesian
coordinate system which is not aligned to the component of the magnetic
field described by the shear.
The standard ``four field model'' is an isothermal two fluid version of
the equations keeping parallel velocities and densities.  In fluid
language it is \cite{recon}
\begin{eqnarray}
\dtt{\Omega} &=& \dpl\Jpl \\
\dtt{n_e} &=& \dpl(\Jpl-\upl) \\
\mu_i\dtt{\upl} &=& -\dpl n_e \\
\beta_e\ptt{\Apl}+\mu_e\dtt{\Jpl} &=& \dpl (n_e-\phi)
\end{eqnarray}
where the vorticity and current are given by
\begin{equation}
\Omega = \ddpp\phi  \qqquad \Jpl = -\ddpp\Apl
\end{equation}
and the nonlinear derivatives are given by
\begin{equation}
\dtt{}=\ptt{}+[\phi,\,] \qqquad \dpl=-\beta_e[\psi,\,]
\end{equation}
This is obviously related to the above gyrofluid model, with cold ions
and with isothermal electrons,
and no FLR effects.  The simplified ``two-field model'' is then
prescribed by neglecting $\upl$ and setting $n_e=\Omega$.  The four
field model is equivalent to the two-moment simplification of the
gyrofluid model, with $\Jpl=\upl-\vpl$ and $\Omega=n_e-n_i$, so that
\begin{eqnarray}
\dtt{n_e} &=& -\dpl\vpl\\
\dtt{n_i} &=& -\dpl\upl\\
\beta_e\ptt{\Apl}+\mu_i\dtt{\upl} &=& -\dpl \phi \\
\beta_e\ptt{\Apl}-\mu_e\dtt{\vpl} &=& -\dpl(\phi-n_e) \\
\end{eqnarray}
up to some $\mu_e/\mu_i$ corrections.  Putting the FLR effects back in
(with constant background temperatures) we then have the two moment
equations for each species,
\begin{eqnarray}
\dtt{\nzfl} &=& -\dpl\uzfl\\
\beta_e\ptt{\psig}+\rmuz\dtt{\uzfl} &=& -\dpl (\phig+\tauz\nzfl) \\
\end{eqnarray}
with nonlinear derivatives,
\begin{equation}
\dtt{}=\ptt{}+[\phig,\,] \qqquad \dpl=-\beta_e[\psig,\,]
\end{equation}
and FLR potentials,
\begin{equation}
\phig=\Gamma_1\phi \qqquad \psig=\Gamma_1\Apl
\end{equation}
and with polarisation,
\begin{equation}
\sumsp\azz\LB\Gamma_1\nzfl+{\Gamma_0-1\over\tauz}\phi\RB = 0
\end{equation}
and induction (cf.\ Eqs.\ \ref{eqpsi},\ref{eqpsig}),
\begin{equation}
\LP\sum_z\azz\LB {\beta_e\over\rmuz}\Gamma_1^2\RB - \ddpp\RP\Apl 
	= \sum_z{\azz\over\rmuz}\LB\Gamma_1\LP\beta_e\psig+\rmuz\uzfl\RP\RB
\end{equation}
Hence, we find that the four field model has this obvious generalised
FLR version, through the two gyrofluid moments, and the two-field model
is basically the same thing with the ion gyrofluid moment variables
neglected on the basis that self consistent gradient drive is absent and
$\beta_e/\mu_i\ll 1$.  The foregoing correspondence
was shown in Refs.\ \cite{eps03,GEM}, in the context of tokamak
microturbulence. 

The obvious next step for reconnection with finite electron gyroradius
modifications to the nonlinear magnetic field dynamics 
is to restore all six gyrofluid moment variables,
\begin{equation}
\ptt{\nzfl} + [\phig,\nzfl] + [\vorg,\tzppfl]
	= \beta_e[\psig,\uzfl] + \beta_e[\chig,\qzppfl]
\end{equation}
\begin{eqnarray}
&& \beta_e\ptt{\psig}+\rmuz\ptt{\uzfl} 
	+ \rmuz[\phig,\uzfl] + \rmuz[\vorg,\qzppfl]
\nonumber\\ && \qquad{}
	= \beta_e[\psig,(\phig+\tauz\pzplfl)]
	+ \beta_e[\chig,(\vorg+\tauz\tzppfl)]
\end{eqnarray}
\begin{equation}
\half\ptt{\tzplfl} + \half[\phig,\tzplfl]
	= \beta_e[\psig,(\uzfl+\qzplfl)] + \beta_e[\chig,\qzppfl]
\end{equation}
\begin{equation}
\ptt{\tzppfl} + [\phig,\tzppfl] + [\vorg,(\nzfl+2\tzppfl)]
	= \beta_e[\psig,\qzppfl] + \beta_e[\chig,(\uzfl+2\qzppfl)]
\end{equation}
\begin{equation}
\rmuz\ptt{\qzplfl} + \rmuz[\phig,\qzplfl]
	= \threehalves\tauz[\psig,\tzplfl]
\end{equation}
\begin{eqnarray}
&& \beta_e\ptt{\chig}+\rmuz\ptt{\qzppfl} 
	+ \rmuz[\phig,\qzppfl]
	+ \rmuz[\vorg,(\uzfl+2\qzppfl)]
\nonumber\\ && \qquad{}
	= \tauz\beta_e[\psig,\tzppfl]
		+ \beta_e[\chig,(\phig+\tauz\pzplfl)] 
		+ \beta_e[\chig,2(\vorg+\tauz\tzppfl)]
\end{eqnarray}
with FLR potentials and polarisation and induction equations as in the
full six-moment gyrofluid model (Eqs.\ \ref{eqphig}--\ref{eqpsi})
above.

\subsection{Constraints on magnetic geometry}

At this level the derivation of a set of equations actually used in a
computation depends on assumptions made about the geometry.  For 2D
reconnection studies in a model containing a guide field the typical
case is that a background homogeneous magnetic field $\vec B_0$
is prescribed, the
coordinates of the computational domain describe the plane perpendicular
to $\vec B_0$, and the components of the magnetic field in this plane,
labelled $xy$, are determined by the evolving potential $\Apl$.  For 3D
magnetised plasma turbulence studies the coordinates are usually aligned
to a background magnetic field with both curvature and shear.  The
derivation of gyrofluid equations does not depend on the use of either,
or other, sort of model.  However, the most important point is this:
energetic consistency must be maintained.  In this context the only
requirement is that the tensor $\vec F_0$ defining the nonlinear
brackets is antisymmetric and divergence free, 
and the vector $\kappacv^i$ defining the curvature operator is
divergence free.  Although the equations were derived using strict
fluxtube ordering, that was made necessary by the dependence of $F^M$
upon $B$ and therefore its spatial dependence.  That was simply an
energetic consistency constraint.
Once the equations are derived, one need only maintain that same level
of energetic consistency, and this is done by retaining the required
properties for $\vec F_0$ and $\kappacv$.  This makes a global geometry
model possible.  Local and global field aligned geometry is given in
Refs.\ \cite{Dewar83,Beergeom} and the necessary constraints of global
consistency of fluxtubes and the transforms necessary to obtain global
mode structure are given in Refs.\ \cite{fluxtube} and \cite{shifted},
respectively.  An example of a global tokamak
geometry model which relaxes the
ordering on the coordinate derivatives, while keeping it on the
equations themselves, is to recast the curvature operator in terms of a
bracket,
\begin{equation}
\kappacv(f) = 2[\log R,f]
\end{equation}
with $\log R$ a scalar function of the coordinates, and the brackets
given as in Eq.\ \ref{eqbracket} with $\log R$ and the components of
$\vec F_0$, along with the metric and magnetic field component $B^s$,
describing the geometry.  
A version of this is given in Ref.\
\cite{shifted} and is suitable for global computation of the tokamak
core, or the edge region within one scale-length variation of the
background parameters dependent upon densities and temperatures.  The
ordering behind the equations essentially limits validity to a domain
comprising one such set of scale lengths.  Otherwise, the parameter set
ceases to be representative (cf.\ the discussion in Refs.\ \cite{eps03},
\cite{eps06}, or \cite{krakow}).
These steps are what extend the GEM model of
Refs.\ \cite{aps99,eps03,GEM} to the GEMR model described
in Refs.\ \cite{pet05,Falchetto08}.

\subsection{Global computation and the issue of stratification}

Global computations with this delta-f gyrofluid model and its
predecessor in Ref.\ \cite{GEM}\ 
are not only
possible but have been underway for some time, both for edge
\cite{pet05,zweben09} and for core \cite{Falchetto08} cases.  With the
background current profile added to the electron parallel velocity
dependent variable 
$\uepl\to\uepl - J_0/(n_e e)$, where $J_0=J_0(x)$, with 
$n_e e$ a normalising
parameter, the model is being used for studies of the self consistent
interaction between dominantly ion temperature gradient driven
electromagnetic core turbulence with magnetic islands \cite{eps10}.  One
often thinks of fluxtube computations as being defined by periodic
radial boundary conditions.  In fact they are defined by taking the
radially local approximation on the flux surface geometry, so that all
metric quantities depend on $s$ only except for the appearance of shear,
and then assuming that $\ppx{},\ppy{}\gg\pps{}$ in the derivatives.
(cf.\ Sec.\ \ref{sec:deltaf}).
Fluxtube computations without periodic boundary conditions are common
(\eg,
Refs.\ \cite{dalfloc,aps99,eps03,Naulin03,Naulin05,zfenergy,krakow}).
Although the delta-f gyrofluid equations are initially derived using
fluxtube approximations on the gyrokinetic model
, in the resulting gyrofluid model
global geometry may be restored as long as the model drift tensor used
to define the brackets is divergence free (cf.\ the previous subsection;
this step cannot be taken directly
on the delta-f gyrokinetic model, due to the
dependence of $F^M$ upon $B$ and the need to commute $F^M$ past
derivatives to conserve free energy).

The other issue faced by truly global computations is stratification.
Although delta-f models can use global geometry and do global
simulations, what they cannot represent is a change in physical
parameters across the domain.  Again, this is for energy conservation
reasons: the physical parameters in the delta-f equations (including
those in $F^M$) must also commute with the nonlinear
bracket operations, including the $\kappacv$ terms.  A stratified
nonlinear model will require that all locally varying parameters (\ie,
temperature, beta) be determined (three-dimensionally) by the dependent
variables.  Then, the conserved quantity is no longer free energy but
the Noether energy, as explained in Sec.\ \ref{sec:energy}.  For a
gyrofluid model to be able to represent stratification, analogous to
nonlocal fluid Braginskii equations 
of the sort given in Ref.\ \cite{transport}, it must either be derived
from a total-f gyrokinetic model such as given in
Sec.\ \ref{sec:totalf}, or the simplified version without FLR effects
used to treat equilibrium flows in Ref.\ \cite{pet09}, or directly from
a model Lagrangian as in Refs.\ \cite{Strintzi04,Strintzi05}.  This is
still ongoing work as the need to have parallel heat fluxes as dynamical
variables and the simultaneous incorporation of FLR effects
requires extension of Refs.\ \cite{Strintzi04,Strintzi05}, which is the
closest at present towards a complete total-f gyrofluid model.

With such a stratified model one can anticipate to capture the spatial
transition between regions of different physical character and perhaps
also temporal transitions.  Of course, to treat such phenomena one
requires a solid foundation within energetic consistency.

\subsection{Relation to dissipative fluid drift equations and Reduced
  MHD
\label{sec:braginskii}}

A detailed analysis of these dissipation terms under the Braginskii
limits (small gyroradius, strong collisionality, and an implicit
assumption that specific heat fluxes are small compared to fluid
velocities) showed that they recover the forms used in reduced (low
frequency, neglecting compressional Alfv\'en dynamics and expressing
perpendicular flows as drifts rather than directly as dependent
variables) Braginskii equations.  Polarisation heat flux effects due 
to finite perpendicular inertia were also recovered.
By extension, the reduced magnetohydrodynamic (MHD) 
equations are already well
known to be a subset of the reduced Braginskii equations.
These results were obtained in a previous work \cite{braggem},
and are summarised here.

The Braginskii regime is defined by the assumptions used by Braginskii
to obtain his collisional fluid equations \cite{Braginskii}, whose low
frequency regime still represents the most commonly used model for
nonlinear edge turbulence computations.  The assumptions are: 
(1) dominance by collisions, such that both $\nu_z\gg\ppt{}$ and
$\nu_z\gg V_z\dpl{}$, where $\nu_z$ is the like
particle collision frequency and $V_z$ is the thermal velocity,
(2) long wavelength, such that $\rho_z^2\ddpp\ll 1$, where $\rho_z$ is
the thermal gyroradius, and, much less commonly understood,
(3) small specific heat flux, such that $\vec q_z\ll p_z\vec u_z$,
implicitly assumed in the use of a drifting Maxwellian to lowest order
and obtaining $\vec q$ only through first order corrections.  In these
inequalities the subscript $z$ denotes the species, \ie, the
inequalities should hold for all species.  The most important thing to
know about tokamak edge turbulence (in particular) is that all three of
these assumptions are violated, and the problems are especially severe
in the ions, not the electrons which received most of the early
attention \cite{Hassam80}.  Specifically, $\nu_i$ is one to two orders
of magnitude slower than the turbulence, for the longer and shorter
wavelength component, respectively, and the turbulence vorticity
spectrum always extends past $\kpp\rs=1$ for drift wave turbulence (in a
plasma with $T_i\sim T_e$ this means a component with $\kpp\rho_i\sim 1$
must be faced), and especially for temperature gradient driven
turbulence, not only are specific parallel heat fluxes larger than
velocities in fluctuations, but the diamagnetic compressional effects in
tokamak geometry scale with the gradients and therefore the heat flux
effects are stronger in both the diamagnetic and polarisation effects in
the ions.  The latter indicates extending the fluid
models to treat heat fluxes as well as velocities in the stress tensor
effects that represent nonlinear polarisation, as introduced and
discussed in Ref.\ \cite{Pogutse98}.

The gyrofluid equations represent all these extension effects as well as
the basic Braginskii ones automatically.  Briefly, nonlinear finite
gyroradius effects in the gyrofluid model reduce term by term to the
nonlinear polarisation velocity and heat flux effects in an expansion
keeping lowest order $\kkpp\rho_z^2$ effects in the ExB advection and in
the equivalence of representation implied by the polarisation equation.
For an isothermal model, one can start with the definition of $\phig$,
solve the polarisation equation for $n_i$ in terms of $n_e$, expand
these expressions one order in $\kkpp\rho_z^2$ and
insert them into the nonlinear advection derivative for $n_i$, and do
some operations involving converting $\ddpp$ inside brackets into
general divergences.  The result is the fluid density equation with the
nonlinear polarisation terms.  With the temperatures the procedure is
more complicated and involves manipulation of the $\vorg$ terms and
derivation of the space/gyrofluid temperature representation from the
moment hierarchy.  This procedure recovers the temperature gradient
parts of the polarisation velocity and also the polarisation
heat flux terms.

The other correspondences involve collisional dissipation formulae and
are simpler.  The parallel viscosity is proportional to the temperature
anisotropy $\delta T_z=\tzplfl-\tzppfl$.  The two temperature equations are
used to form a time dependent equation for $\delta T_z$.  The velocity
divergences ($\dpl$ and $\kappacv$ terms) and collisional dissipation
are put on the right side, and the time derivatives, nonlinearities, and
heat flux divergences are put on the left side.  The Braginskii regime
consists of assuming the left side to be small, recovering the
collisional formula for parallel viscosity (since we set the $\pi_z$
coefficients accordingly).  As noted, this is severely violated for ions
in edge turbulence.  This result was also obtained in Ref.\ \cite{pet05}.

The parallel heat flux in a collisional model is the sum of the
perp-parallel and parallel-parallel components.  The equation
for these are added, the temperature gradients and collisional
damping terms are put on the right side and all other terms are put 
on the left side.  The Braginskii regime
consists of assuming the left side to be small, recovering the
collisional formula for parallel heat flux (for this we set the
$\kappa_z$ coefficients accordingly).  This is also violated for ions,
with the ExB advection nonlinearities much larger than any of the
dissipation terms for typical parameters.
This result was also obtained in Ref.\ \cite{pet05}.

Once the Braginskii regime is (formally) recovered, the next steps to
reduced MHD are well known (this is a matter of the difference between
two-fluid models, \eg, Wakatani-Hasegawa \cite{HasWak83,WakHas84}, and the
one-fluid versions of reduced MHD \cite{Strauss76,Strauss77}).  At the
isothermal level, leading to what are called four-field models, this was
done in detail with both models and results for edge turbulence in
Ref.\ \cite{bdmode}.  The basic assumptions are that $e\phifl/T_e$ is
larger than any relative fluctuation variable in densities or
temperatures, and that compressibility effects in $\Jpl$ are neglected
(energetically, the second assumption follows from the first).  This can
be called MHD ordering (pressure gradients are neglected in favour of
electric field components).

At the philosophical level, the gyrofluid model treats polarisation
densities instead of velocities, but the simplest road back to reduced
MHD is to apply $\ppt{}$ to the polarisation equation, apply MHD
ordering, and follow all the consequences.  Similarly, the Ohm's law
amounts to neglecting electron inertia effects and applying MHD ordering
to the electron parallel velocity equation.  Reduced MHD without
polarisation effects is particularly easy to understand (neglect all FLR
effects, subtract the continuity equations, and replace the gyrocenter
charge density with the ExB vorticity using the polarisation effects).
A more introductory version of this correspondence between models was
given in Ref.\ \cite{eps03}.  Basically, the gyrofluid model solves the
same problems as the reduced Braginskii fluid model, without the more
damaging assumptions of the latter.

\section{Summary and comment}

The gyrofluid model corresponding to what is known as the ``delta-f''
gyrokinetic theory now has a derivation path which starts from first
principles.  It depends on the ordering used to obtain the delta-f forms
of both the gyrokinetic equation and the associated field potential
equations.  It further depends on a list of moment variables, the
description of the underlying distribution function in terms of these
variables, and essentially one assumption involving the moment closure
of finite gyroradius (FLR) corrections.  The energy theorem descends
from both the total- and delta-f gyrokinetic versions and is used to fix
the rest of the undetermined quantities in the gyrofluid model.  The
resulting gyrofluid model is now consistent at the level of the best
delta-f gyrokinetic models, especially in terms of an energy theorem
which is properly conservative in all the reactive effects
(compressibility, coupling to flows and MHD, etc.), and also in terms of
the responses of heat fluxes to temperature gradients being carried at
the same level of sophistication as the responses of flows and currents
to density gradients and electric fields.  For dynamics of magnetised
plasmas driven principally by temperature gradients this is the minimal
fluid model.  For dynamics more generically pressure or current driven
the isothermal version \cite{eps03}, easily and consistently obtained by
setting the temperature and heat flux moment variables, the second FLR
operation $\Gamma_2$, and the thermoelectric collisional effect (through
$\alpha_e$) to zero, becomes the minimal model.  Hence, the decision of
what constitutes the minimal model depends on the problem being
considered.  However, the procedure given herein is useful generally in
constraining the derivation of any gyrofluid model based upon, as it
must be, an underlying gyrokinetic model which is energetically
consistent.

{
\bibliography{../paper}

\begin{thebibliography}{72}
\expandafter\ifx\csname natexlab\endcsname\relax\def\natexlab#1{#1}\fi
\expandafter\ifx\csname bibnamefont\endcsname\relax
  \def\bibnamefont#1{#1}\fi
\expandafter\ifx\csname bibfnamefont\endcsname\relax
  \def\bibfnamefont#1{#1}\fi
\expandafter\ifx\csname citenamefont\endcsname\relax
  \def\citenamefont#1{#1}\fi
\expandafter\ifx\csname url\endcsname\relax
  \def\url#1{\texttt{#1}}\fi
\expandafter\ifx\csname urlprefix\endcsname\relax\def\urlprefix{URL }\fi
\providecommand{\bibinfo}[2]{#2}
\providecommand{\eprint}[2][]{\url{#2}}

\bibitem[{\citenamefont{Beer and Hammett}(1996)}]{Beer96}
\bibinfo{author}{\bibfnamefont{M.~A.} \bibnamefont{Beer}} \bibnamefont{and}
  \bibinfo{author}{\bibfnamefont{G.}~\bibnamefont{Hammett}},
  \bibinfo{journal}{Phys. Plasmas} \textbf{\bibinfo{volume}{3}},
  \bibinfo{pages}{4046} (\bibinfo{year}{1996}).

\bibitem[{\citenamefont{Dorland and Hammett}(1993)}]{Dorland93}
\bibinfo{author}{\bibfnamefont{W.}~\bibnamefont{Dorland}} \bibnamefont{and}
  \bibinfo{author}{\bibfnamefont{G.}~\bibnamefont{Hammett}},
  \bibinfo{journal}{Phys. Fluids B} \textbf{\bibinfo{volume}{5}},
  \bibinfo{pages}{812} (\bibinfo{year}{1993}).

\bibitem[{\citenamefont{Hammett and Perkins}(1990)}]{Hammett90}
\bibinfo{author}{\bibfnamefont{G.~W.} \bibnamefont{Hammett}} \bibnamefont{and}
  \bibinfo{author}{\bibfnamefont{F.~W.} \bibnamefont{Perkins}},
  \bibinfo{journal}{Phys. Rev. Lett.} \textbf{\bibinfo{volume}{64}},
  \bibinfo{pages}{3019} (\bibinfo{year}{1990}).

\bibitem[{\citenamefont{Scott}(2000)}]{aps99}
\bibinfo{author}{\bibfnamefont{B.}~\bibnamefont{Scott}},
  \bibinfo{journal}{Phys. Plasmas} \textbf{\bibinfo{volume}{7}},
  \bibinfo{pages}{1845} (\bibinfo{year}{2000}).

\bibitem[{\citenamefont{Snyder and Hammett}(2001)}]{Snyder01}
\bibinfo{author}{\bibfnamefont{P.}~\bibnamefont{Snyder}} \bibnamefont{and}
  \bibinfo{author}{\bibfnamefont{G.}~\bibnamefont{Hammett}},
  \bibinfo{journal}{Phys. Plasmas} \textbf{\bibinfo{volume}{8}},
  \bibinfo{pages}{744} (\bibinfo{year}{2001}).

\bibitem[{\citenamefont{Knorr et~al.}(1988)\citenamefont{Knorr, Hansen, Lynov,
  P\'ecseli, and Rasmussen}}]{Knorr88}
\bibinfo{author}{\bibfnamefont{G.}~\bibnamefont{Knorr}},
  \bibinfo{author}{\bibfnamefont{F.~R.} \bibnamefont{Hansen}},
  \bibinfo{author}{\bibfnamefont{J.~P.} \bibnamefont{Lynov}},
  \bibinfo{author}{\bibfnamefont{H.~L.} \bibnamefont{P\'ecseli}},
  \bibnamefont{and} \bibinfo{author}{\bibfnamefont{J.~J.}
  \bibnamefont{Rasmussen}}, \bibinfo{journal}{Physica Scripta}
  \textbf{\bibinfo{volume}{38}}, \bibinfo{pages}{829} (\bibinfo{year}{1988}).

\bibitem[{\citenamefont{Lee}(1983)}]{Lee83}
\bibinfo{author}{\bibfnamefont{W.~W.} \bibnamefont{Lee}},
  \bibinfo{journal}{Phys. Fluids} \textbf{\bibinfo{volume}{26}},
  \bibinfo{pages}{556} (\bibinfo{year}{1983}).

\bibitem[{\citenamefont{Frieman and Chen}(1982)}]{FriemanChen82}
\bibinfo{author}{\bibfnamefont{E.~A.} \bibnamefont{Frieman}} \bibnamefont{and}
  \bibinfo{author}{\bibfnamefont{L.}~\bibnamefont{Chen}},
  \bibinfo{journal}{Phys. Fluids} \textbf{\bibinfo{volume}{25}},
  \bibinfo{pages}{502} (\bibinfo{year}{1982}).

\bibitem[{\citenamefont{Grad}(1949)}]{Grad}
\bibinfo{author}{\bibfnamefont{H.}~\bibnamefont{Grad}},
  \bibinfo{journal}{Commun. Pure Appl. Math.} \textbf{\bibinfo{volume}{2}},
  \bibinfo{pages}{331} (\bibinfo{year}{1949}).

\bibitem[{\citenamefont{Braginskii}(1965)}]{Braginskii}
\bibinfo{author}{\bibfnamefont{S.~I.} \bibnamefont{Braginskii}},
  \bibinfo{journal}{Rev. Plasma Phys.} \textbf{\bibinfo{volume}{1}},
  \bibinfo{pages}{205} (\bibinfo{year}{1965}).

\bibitem[{\citenamefont{Littlejohn}(1983)}]{Littlejohn83}
\bibinfo{author}{\bibfnamefont{R.}~\bibnamefont{Littlejohn}},
  \bibinfo{journal}{J. Plasma Phys.} \textbf{\bibinfo{volume}{29}},
  \bibinfo{pages}{111} (\bibinfo{year}{1983}).

\bibitem[{\citenamefont{Dubin et~al.}(1983)\citenamefont{Dubin, Krommes,
  Oberman, and Lee}}]{Dubin83}
\bibinfo{author}{\bibfnamefont{D.~H.~E.} \bibnamefont{Dubin}},
  \bibinfo{author}{\bibfnamefont{J.~A.} \bibnamefont{Krommes}},
  \bibinfo{author}{\bibfnamefont{C.}~\bibnamefont{Oberman}}, \bibnamefont{and}
  \bibinfo{author}{\bibfnamefont{W.~W.} \bibnamefont{Lee}},
  \bibinfo{journal}{Phys. Fluids} \textbf{\bibinfo{volume}{26}},
  \bibinfo{pages}{3524} (\bibinfo{year}{1983}).

\bibitem[{\citenamefont{Hahm}(1988)}]{Hahm88}
\bibinfo{author}{\bibfnamefont{T.~S.} \bibnamefont{Hahm}},
  \bibinfo{journal}{Phys. Fluids} \textbf{\bibinfo{volume}{31}},
  \bibinfo{pages}{2670} (\bibinfo{year}{1988}).

\bibitem[{\citenamefont{Scott}(2003{\natexlab{a}})}]{eps03}
\bibinfo{author}{\bibfnamefont{B.}~\bibnamefont{Scott}},
  \bibinfo{journal}{Plasma Phys. Contr. Fusion} \textbf{\bibinfo{volume}{45}},
  \bibinfo{pages}{A385} (\bibinfo{year}{2003}{\natexlab{a}}).

\bibitem[{\citenamefont{Scott}(2005{\natexlab{a}})}]{GEM}
\bibinfo{author}{\bibfnamefont{B.}~\bibnamefont{Scott}},
  \bibinfo{journal}{Phys. Plasmas} \textbf{\bibinfo{volume}{12}},
  \bibinfo{pages}{102307} (\bibinfo{year}{2005}{\natexlab{a}}),
  \eprint{arXiv:physics/0501124}.

\bibitem[{\citenamefont{Sugama}(2000)}]{Sugama00}
\bibinfo{author}{\bibfnamefont{H.}~\bibnamefont{Sugama}},
  \bibinfo{journal}{Phys. Plasmas} \textbf{\bibinfo{volume}{7}},
  \bibinfo{pages}{466} (\bibinfo{year}{2000}).

\bibitem[{\citenamefont{Brizard}(2000)}]{Brizard00}
\bibinfo{author}{\bibfnamefont{A.}~\bibnamefont{Brizard}},
  \bibinfo{journal}{Phys. Plasmas} \textbf{\bibinfo{volume}{7}},
  \bibinfo{pages}{4816} (\bibinfo{year}{2000}).

\bibitem[{\citenamefont{Strintzi and Scott}(2004)}]{Strintzi04}
\bibinfo{author}{\bibfnamefont{D.}~\bibnamefont{Strintzi}} \bibnamefont{and}
  \bibinfo{author}{\bibfnamefont{B.}~\bibnamefont{Scott}},
  \bibinfo{journal}{Phys. Plasmas} \textbf{\bibinfo{volume}{11}},
  \bibinfo{pages}{5452} (\bibinfo{year}{2004}).

\bibitem[{\citenamefont{Strintzi et~al.}(2005)\citenamefont{Strintzi, Scott,
  and Brizard}}]{Strintzi05}
\bibinfo{author}{\bibfnamefont{D.}~\bibnamefont{Strintzi}},
  \bibinfo{author}{\bibfnamefont{B.}~\bibnamefont{Scott}}, \bibnamefont{and}
  \bibinfo{author}{\bibfnamefont{A.}~\bibnamefont{Brizard}},
  \bibinfo{journal}{Phys. Plasmas} \textbf{\bibinfo{volume}{12}},
  \bibinfo{pages}{052517} (\bibinfo{year}{2005}).

\bibitem[{\citenamefont{Lee and Tang}(1988)}]{LeeTang88}
\bibinfo{author}{\bibfnamefont{W.~W.} \bibnamefont{Lee}} \bibnamefont{and}
  \bibinfo{author}{\bibfnamefont{W.~M.} \bibnamefont{Tang}},
  \bibinfo{journal}{Phys. Fluids} \textbf{\bibinfo{volume}{31}},
  \bibinfo{pages}{612} (\bibinfo{year}{1988}).

\bibitem[{\citenamefont{Hasegawa and Mima}(1978)}]{Hasmim78}
\bibinfo{author}{\bibfnamefont{A.}~\bibnamefont{Hasegawa}} \bibnamefont{and}
  \bibinfo{author}{\bibfnamefont{K.}~\bibnamefont{Mima}},
  \bibinfo{journal}{Phys. Fluids} \textbf{\bibinfo{volume}{21}},
  \bibinfo{pages}{87} (\bibinfo{year}{1978}).

\bibitem[{\citenamefont{Montgomery and Turner}(1980)}]{Turner82}
\bibinfo{author}{\bibfnamefont{D.}~\bibnamefont{Montgomery}} \bibnamefont{and}
  \bibinfo{author}{\bibfnamefont{L.}~\bibnamefont{Turner}},
  \bibinfo{journal}{Phys. Fluids} \textbf{\bibinfo{volume}{23}},
  \bibinfo{pages}{264} (\bibinfo{year}{1980}).

\bibitem[{\citenamefont{Wakatani and Hasegawa}(1984)}]{WakHas84}
\bibinfo{author}{\bibfnamefont{M.}~\bibnamefont{Wakatani}} \bibnamefont{and}
  \bibinfo{author}{\bibfnamefont{A.}~\bibnamefont{Hasegawa}},
  \bibinfo{journal}{Phys. Fluids} \textbf{\bibinfo{volume}{27}},
  \bibinfo{pages}{611} (\bibinfo{year}{1984}).

\bibitem[{\citenamefont{Scott}(1992)}]{ssdw}
\bibinfo{author}{\bibfnamefont{B.}~\bibnamefont{Scott}},
  \bibinfo{journal}{Phys. Fluids B} \textbf{\bibinfo{volume}{4}},
  \bibinfo{pages}{2468} (\bibinfo{year}{1992}).

\bibitem[{\citenamefont{Krommes and Hu}(1994)}]{Krommes94}
\bibinfo{author}{\bibfnamefont{J.~A.} \bibnamefont{Krommes}} \bibnamefont{and}
  \bibinfo{author}{\bibfnamefont{G.}~\bibnamefont{Hu}}, \bibinfo{journal}{Phys.
  Plasmas} \textbf{\bibinfo{volume}{1}}, \bibinfo{pages}{3211}
  (\bibinfo{year}{1994}).

\bibitem[{\citenamefont{Sugama et~al.}(2001)\citenamefont{Sugama, Watanabe, and
  Horton}}]{Sugama01}
\bibinfo{author}{\bibfnamefont{H.}~\bibnamefont{Sugama}},
  \bibinfo{author}{\bibfnamefont{T.}~\bibnamefont{Watanabe}}, \bibnamefont{and}
  \bibinfo{author}{\bibfnamefont{W.}~\bibnamefont{Horton}},
  \bibinfo{journal}{Phys. Plasmas} \textbf{\bibinfo{volume}{8}},
  \bibinfo{pages}{2617} (\bibinfo{year}{2001}).

\bibitem[{\citenamefont{Porcelli}(1991)}]{Porcelli91}
\bibinfo{author}{\bibfnamefont{F.}~\bibnamefont{Porcelli}},
  \bibinfo{journal}{Phys. Rev. Lett.} \textbf{\bibinfo{volume}{66}},
  \bibinfo{pages}{425} (\bibinfo{year}{1991}).

\bibitem[{\citenamefont{Schep et~al.}(1994)\citenamefont{Schep, Pegoraro, and
  Kuvshinov}}]{Schep94}
\bibinfo{author}{\bibfnamefont{T.~J.} \bibnamefont{Schep}},
  \bibinfo{author}{\bibfnamefont{F.}~\bibnamefont{Pegoraro}}, \bibnamefont{and}
  \bibinfo{author}{\bibfnamefont{B.~N.} \bibnamefont{Kuvshinov}},
  \bibinfo{journal}{Phys. Plasmas} \textbf{\bibinfo{volume}{1}},
  \bibinfo{pages}{2843} (\bibinfo{year}{1994}).

\bibitem[{\citenamefont{Grasso et~al.}(1999)\citenamefont{Grasso, Pegoraro,
  Porcelli, and Califano}}]{Grasso99}
\bibinfo{author}{\bibfnamefont{D.}~\bibnamefont{Grasso}},
  \bibinfo{author}{\bibfnamefont{F.}~\bibnamefont{Pegoraro}},
  \bibinfo{author}{\bibfnamefont{F.}~\bibnamefont{Porcelli}}, \bibnamefont{and}
  \bibinfo{author}{\bibfnamefont{F.}~\bibnamefont{Califano}},
  \bibinfo{journal}{Plasma Phys. Contr. Fusion} \textbf{\bibinfo{volume}{41}},
  \bibinfo{pages}{1497} (\bibinfo{year}{1999}).

\bibitem[{\citenamefont{Scott and Porcelli}(2004)}]{recon}
\bibinfo{author}{\bibfnamefont{B.}~\bibnamefont{Scott}} \bibnamefont{and}
  \bibinfo{author}{\bibfnamefont{F.}~\bibnamefont{Porcelli}},
  \bibinfo{journal}{Phys. Plasmas} \textbf{\bibinfo{volume}{11}},
  \bibinfo{pages}{5468} (\bibinfo{year}{2004}).

\bibitem[{\citenamefont{Grandgirard et~al.}(2007)\citenamefont{Grandgirard,
  Sarazin, Angelino, Bottino, Crouseilles, Darmet, Dif-Pradalier, Garbet,
  Ghendrih, Jolliet et~al.}}]{Virginie07}
\bibinfo{author}{\bibfnamefont{V.}~\bibnamefont{Grandgirard}},
  \bibinfo{author}{\bibfnamefont{Y.}~\bibnamefont{Sarazin}},
  \bibinfo{author}{\bibfnamefont{P.}~\bibnamefont{Angelino}},
  \bibinfo{author}{\bibfnamefont{A.}~\bibnamefont{Bottino}},
  \bibinfo{author}{\bibfnamefont{N.}~\bibnamefont{Crouseilles}},
  \bibinfo{author}{\bibfnamefont{G.}~\bibnamefont{Darmet}},
  \bibinfo{author}{\bibfnamefont{G.}~\bibnamefont{Dif-Pradalier}},
  \bibinfo{author}{\bibfnamefont{X.}~\bibnamefont{Garbet}},
  \bibinfo{author}{\bibfnamefont{P.}~\bibnamefont{Ghendrih}},
  \bibinfo{author}{\bibfnamefont{S.}~\bibnamefont{Jolliet}},
  \bibnamefont{et~al.}, \bibinfo{journal}{Plasma Phys. Contr. Fusion}
  \textbf{\bibinfo{volume}{49}}, \bibinfo{pages}{B173} (\bibinfo{year}{2007}).

\bibitem[{\citenamefont{Idomura et~al.}(2007)\citenamefont{Idomura, Ida,
  Tokuda, and Villard}}]{Idomura07}
\bibinfo{author}{\bibfnamefont{Y.}~\bibnamefont{Idomura}},
  \bibinfo{author}{\bibfnamefont{M.}~\bibnamefont{Ida}},
  \bibinfo{author}{\bibfnamefont{S.}~\bibnamefont{Tokuda}}, \bibnamefont{and}
  \bibinfo{author}{\bibfnamefont{L.}~\bibnamefont{Villard}},
  \bibinfo{journal}{J. Comput. Phys.} \textbf{\bibinfo{volume}{226}},
  \bibinfo{pages}{244} (\bibinfo{year}{2007}).

\bibitem[{\citenamefont{Garbet et~al.}(2007)\citenamefont{Garbet, Sarazin,
  Grandgirard, Dif-Pradalier, Darmet, Ghendrih, Angelino, Bertrand, Besse,
  Gravier et~al.}}]{Garbet07}
\bibinfo{author}{\bibfnamefont{X.}~\bibnamefont{Garbet}},
  \bibinfo{author}{\bibfnamefont{Y.}~\bibnamefont{Sarazin}},
  \bibinfo{author}{\bibfnamefont{V.}~\bibnamefont{Grandgirard}},
  \bibinfo{author}{\bibfnamefont{G.}~\bibnamefont{Dif-Pradalier}},
  \bibinfo{author}{\bibfnamefont{G.}~\bibnamefont{Darmet}},
  \bibinfo{author}{\bibfnamefont{P.}~\bibnamefont{Ghendrih}},
  \bibinfo{author}{\bibfnamefont{P.}~\bibnamefont{Angelino}},
  \bibinfo{author}{\bibfnamefont{P.}~\bibnamefont{Bertrand}},
  \bibinfo{author}{\bibfnamefont{N.}~\bibnamefont{Besse}},
  \bibinfo{author}{\bibfnamefont{E.}~\bibnamefont{Gravier}},
  \bibnamefont{et~al.}, \bibinfo{journal}{Nucl. Fusion}
  \textbf{\bibinfo{volume}{47}}, \bibinfo{pages}{1206} (\bibinfo{year}{2007}).

\bibitem[{\citenamefont{Hahm et~al.}(1988)\citenamefont{Hahm, Lee, and
  Brizard}}]{Hahm88a}
\bibinfo{author}{\bibfnamefont{T.~S.} \bibnamefont{Hahm}},
  \bibinfo{author}{\bibfnamefont{W.~W.} \bibnamefont{Lee}}, \bibnamefont{and}
  \bibinfo{author}{\bibfnamefont{A.}~\bibnamefont{Brizard}},
  \bibinfo{journal}{Phys. Fluids} \textbf{\bibinfo{volume}{31}},
  \bibinfo{pages}{1940} (\bibinfo{year}{1988}).

\bibitem[{\citenamefont{Beer et~al.}(1995)\citenamefont{Beer, Cowley, and
  Hammett}}]{Beergeom}
\bibinfo{author}{\bibfnamefont{M.}~\bibnamefont{Beer}},
  \bibinfo{author}{\bibfnamefont{S.}~\bibnamefont{Cowley}}, \bibnamefont{and}
  \bibinfo{author}{\bibfnamefont{G.}~\bibnamefont{Hammett}},
  \bibinfo{journal}{Phys. Plasmas} \textbf{\bibinfo{volume}{2}},
  \bibinfo{pages}{2687} (\bibinfo{year}{1995}).

\bibitem[{\citenamefont{Scott}(1998)}]{fluxtube}
\bibinfo{author}{\bibfnamefont{B.}~\bibnamefont{Scott}},
  \bibinfo{journal}{Phys. Plasmas} \textbf{\bibinfo{volume}{5}},
  \bibinfo{pages}{2334} (\bibinfo{year}{1998}).

\bibitem[{\citenamefont{Scott}(2001)}]{shifted}
\bibinfo{author}{\bibfnamefont{B.}~\bibnamefont{Scott}},
  \bibinfo{journal}{Phys. Plasmas} \textbf{\bibinfo{volume}{8}},
  \bibinfo{pages}{447} (\bibinfo{year}{2001}).

\bibitem[{\citenamefont{Gagne and Shoucri}(1977)}]{Shoucri77}
\bibinfo{author}{\bibfnamefont{R.}~\bibnamefont{Gagne}} \bibnamefont{and}
  \bibinfo{author}{\bibfnamefont{M.~M.} \bibnamefont{Shoucri}},
  \bibinfo{journal}{J. Comput. Phys.} \textbf{\bibinfo{volume}{24}},
  \bibinfo{pages}{445} (\bibinfo{year}{1977}).

\bibitem[{\citenamefont{Shoucri}(1978)}]{Shoucri78}
\bibinfo{author}{\bibfnamefont{M.~M.} \bibnamefont{Shoucri}},
  \bibinfo{journal}{Phys. Fluids} \textbf{\bibinfo{volume}{21}},
  \bibinfo{pages}{1359} (\bibinfo{year}{1978}).

\bibitem[{\citenamefont{Ghizzo et~al.}(1988)\citenamefont{Ghizzo, Izrar,
  Bertrand, Fijalkow, Feix, and Shoucri}}]{Ghizzo}
\bibinfo{author}{\bibfnamefont{A.}~\bibnamefont{Ghizzo}},
  \bibinfo{author}{\bibfnamefont{B.}~\bibnamefont{Izrar}},
  \bibinfo{author}{\bibfnamefont{P.}~\bibnamefont{Bertrand}},
  \bibinfo{author}{\bibfnamefont{E.}~\bibnamefont{Fijalkow}},
  \bibinfo{author}{\bibfnamefont{M.~R.} \bibnamefont{Feix}}, \bibnamefont{and}
  \bibinfo{author}{\bibfnamefont{M.}~\bibnamefont{Shoucri}},
  \bibinfo{journal}{Phys. Fluids} \textbf{\bibinfo{volume}{31}},
  \bibinfo{pages}{72} (\bibinfo{year}{1988}).

\bibitem[{\citenamefont{Bertrand et~al.}(1990)\citenamefont{Bertrand, Ghizzo,
  Johnston, Shoucri, Fijalkow, and Feix}}]{Bertrand90}
\bibinfo{author}{\bibfnamefont{P.}~\bibnamefont{Bertrand}},
  \bibinfo{author}{\bibfnamefont{A.}~\bibnamefont{Ghizzo}},
  \bibinfo{author}{\bibfnamefont{T.~W.} \bibnamefont{Johnston}},
  \bibinfo{author}{\bibfnamefont{M.}~\bibnamefont{Shoucri}},
  \bibinfo{author}{\bibfnamefont{E.}~\bibnamefont{Fijalkow}}, \bibnamefont{and}
  \bibinfo{author}{\bibfnamefont{M.~R.} \bibnamefont{Feix}},
  \bibinfo{journal}{Phys. Fluids B} \textbf{\bibinfo{volume}{2}},
  \bibinfo{pages}{1028} (\bibinfo{year}{1990}).

\bibitem[{\citenamefont{Dannert and Jenko}(2004)}]{Dannert04}
\bibinfo{author}{\bibfnamefont{T.}~\bibnamefont{Dannert}} \bibnamefont{and}
  \bibinfo{author}{\bibfnamefont{F.}~\bibnamefont{Jenko}},
  \bibinfo{journal}{Comput. Phys. Comm.} \textbf{\bibinfo{volume}{163}},
  \bibinfo{pages}{67} (\bibinfo{year}{2004}).

\bibitem[{\citenamefont{Dannert and Jenko}(2005)}]{Dannert05}
\bibinfo{author}{\bibfnamefont{T.}~\bibnamefont{Dannert}} \bibnamefont{and}
  \bibinfo{author}{\bibfnamefont{F.}~\bibnamefont{Jenko}},
  \bibinfo{journal}{Phys. Plasmas} \textbf{\bibinfo{volume}{12}},
  \bibinfo{pages}{07239} (\bibinfo{year}{2005}).

\bibitem[{\citenamefont{Brizard}(2004)}]{Brizard04}
\bibinfo{author}{\bibfnamefont{A.}~\bibnamefont{Brizard}},
  \bibinfo{journal}{Phys. Plasmas} \textbf{\bibinfo{volume}{11}},
  \bibinfo{pages}{4429} (\bibinfo{year}{2004}).

\bibitem[{\citenamefont{Mishchenko and K{\"o}nies}(2007)}]{Mishchenko07}
\bibinfo{author}{\bibfnamefont{A.}~\bibnamefont{Mishchenko}} \bibnamefont{and}
  \bibinfo{author}{\bibfnamefont{A.}~\bibnamefont{K{\"o}nies}},
  \bibinfo{journal}{J. Plasma Phys.} \textbf{\bibinfo{volume}{73}},
  \bibinfo{pages}{757} (\bibinfo{year}{2007}).

\bibitem[{\citenamefont{Scott}(2006{\natexlab{a}})}]{hmode05}
\bibinfo{author}{\bibfnamefont{B.}~\bibnamefont{Scott}},
  \bibinfo{journal}{Plasma Phys. Contr. Fusion} \textbf{\bibinfo{volume}{48}},
  \bibinfo{pages}{A387} (\bibinfo{year}{2006}{\natexlab{a}}).

\bibitem[{\citenamefont{Scott}(1997)}]{dalfloc}
\bibinfo{author}{\bibfnamefont{B.}~\bibnamefont{Scott}},
  \bibinfo{journal}{Plasma Phys. Contr. Fusion} \textbf{\bibinfo{volume}{39}},
  \bibinfo{pages}{1635} (\bibinfo{year}{1997}).

\bibitem[{\citenamefont{Xu and Cohen}(1998)}]{Xu97}
\bibinfo{author}{\bibfnamefont{X.~Q.} \bibnamefont{Xu}} \bibnamefont{and}
  \bibinfo{author}{\bibfnamefont{R.~H.} \bibnamefont{Cohen}},
  \bibinfo{journal}{Contrib. Plasma Phys.} \textbf{\bibinfo{volume}{38}},
  \bibinfo{pages}{158} (\bibinfo{year}{1998}).

\bibitem[{\citenamefont{Rogers et~al.}(1998)\citenamefont{Rogers, Drake, and
  Zeiler}}]{Rogers98}
\bibinfo{author}{\bibfnamefont{B.~N.} \bibnamefont{Rogers}},
  \bibinfo{author}{\bibfnamefont{J.~F.} \bibnamefont{Drake}}, \bibnamefont{and}
  \bibinfo{author}{\bibfnamefont{A.}~\bibnamefont{Zeiler}},
  \bibinfo{journal}{Phys. Rev. Lett.} \textbf{\bibinfo{volume}{81}},
  \bibinfo{pages}{4396} (\bibinfo{year}{1998}).

\bibitem[{\citenamefont{Hassam}(1980)}]{Hassam80}
\bibinfo{author}{\bibfnamefont{A.~B.} \bibnamefont{Hassam}},
  \bibinfo{journal}{Phys. Fluids} \textbf{\bibinfo{volume}{23}},
  \bibinfo{pages}{38} (\bibinfo{year}{1980}).

\bibitem[{\citenamefont{Scott}(2006{\natexlab{b}})}]{pet05}
\bibinfo{author}{\bibfnamefont{B.}~\bibnamefont{Scott}},
  \bibinfo{journal}{Contrib. Plasma Phys.} \textbf{\bibinfo{volume}{46}},
  \bibinfo{pages}{714} (\bibinfo{year}{2006}{\natexlab{b}}).

\bibitem[{\citenamefont{Ottaviani and Porcelli}(1993)}]{Ottaviani93}
\bibinfo{author}{\bibfnamefont{M.}~\bibnamefont{Ottaviani}} \bibnamefont{and}
  \bibinfo{author}{\bibfnamefont{F.}~\bibnamefont{Porcelli}},
  \bibinfo{journal}{Phys. Rev. Lett.} \textbf{\bibinfo{volume}{71}},
  \bibinfo{pages}{3802} (\bibinfo{year}{1993}).

\bibitem[{\citenamefont{Ottaviani and Porcelli}(1995)}]{Ottaviani95}
\bibinfo{author}{\bibfnamefont{M.}~\bibnamefont{Ottaviani}} \bibnamefont{and}
  \bibinfo{author}{\bibfnamefont{F.}~\bibnamefont{Porcelli}},
  \bibinfo{journal}{Phys. Plasmas} \textbf{\bibinfo{volume}{2}},
  \bibinfo{pages}{4104} (\bibinfo{year}{1995}).

\bibitem[{\citenamefont{Fitzpatrick and Porcelli}(2004)}]{Fitzpatrick04}
\bibinfo{author}{\bibfnamefont{R.}~\bibnamefont{Fitzpatrick}} \bibnamefont{and}
  \bibinfo{author}{\bibfnamefont{F.}~\bibnamefont{Porcelli}},
  \bibinfo{journal}{Phys. Plasmas} \textbf{\bibinfo{volume}{11}},
  \bibinfo{pages}{4713} (\bibinfo{year}{2004}).

\bibitem[{\citenamefont{Fitzpatrick and Porcelli}(2007)}]{Fitzpatrick07}
\bibinfo{author}{\bibfnamefont{R.}~\bibnamefont{Fitzpatrick}} \bibnamefont{and}
  \bibinfo{author}{\bibfnamefont{F.}~\bibnamefont{Porcelli}},
  \bibinfo{journal}{Phys. Plasmas} \textbf{\bibinfo{volume}{14}},
  \bibinfo{pages}{049902} (\bibinfo{year}{2007}).

\bibitem[{\citenamefont{Dewar and Glasser}(1983)}]{Dewar83}
\bibinfo{author}{\bibfnamefont{R.~L.} \bibnamefont{Dewar}} \bibnamefont{and}
  \bibinfo{author}{\bibfnamefont{A.~H.} \bibnamefont{Glasser}},
  \bibinfo{journal}{Phys. Fluids} \textbf{\bibinfo{volume}{26}},
  \bibinfo{pages}{3038} (\bibinfo{year}{1983}).

\bibitem[{\citenamefont{Scott}(2006{\natexlab{c}})}]{eps06}
\bibinfo{author}{\bibfnamefont{B.}~\bibnamefont{Scott}},
  \bibinfo{journal}{Plasma Phys. Contr. Fusion} \textbf{\bibinfo{volume}{48}},
  \bibinfo{pages}{B277} (\bibinfo{year}{2006}{\natexlab{c}}).

\bibitem[{\citenamefont{Scott}(2007{\natexlab{a}})}]{krakow}
\bibinfo{author}{\bibfnamefont{B.}~\bibnamefont{Scott}},
  \bibinfo{journal}{Plasma Phys. Contr. Fusion} \textbf{\bibinfo{volume}{49}},
  \bibinfo{pages}{S25} (\bibinfo{year}{2007}{\natexlab{a}}).

\bibitem[{\citenamefont{Falchetto et~al.}(2008)\citenamefont{Falchetto, Scott,
  Angelino, Bottino, Dannert, Grandgirard, Janhunen, Jenko, Jolliet, Kendl
  et~al.}}]{Falchetto08}
\bibinfo{author}{\bibfnamefont{G.~L.} \bibnamefont{Falchetto}},
  \bibinfo{author}{\bibfnamefont{B.~D.} \bibnamefont{Scott}},
  \bibinfo{author}{\bibfnamefont{P.}~\bibnamefont{Angelino}},
  \bibinfo{author}{\bibfnamefont{A.}~\bibnamefont{Bottino}},
  \bibinfo{author}{\bibfnamefont{T.}~\bibnamefont{Dannert}},
  \bibinfo{author}{\bibfnamefont{V.}~\bibnamefont{Grandgirard}},
  \bibinfo{author}{\bibfnamefont{S.~J.} \bibnamefont{Janhunen}},
  \bibinfo{author}{\bibfnamefont{F.}~\bibnamefont{Jenko}},
  \bibinfo{author}{\bibfnamefont{S.}~\bibnamefont{Jolliet}},
  \bibinfo{author}{\bibfnamefont{A.}~\bibnamefont{Kendl}},
  \bibnamefont{et~al.}, \bibinfo{journal}{Plasma Phys. Contr. Fusion}
  \textbf{\bibinfo{volume}{50}}, \bibinfo{pages}{124015}
  (\bibinfo{year}{2008}).

\bibitem[{\citenamefont{Zweben et~al.}(2009)\citenamefont{Zweben, Scott, Terry,
  LaBombard, Hughes, and Stotler}}]{zweben09}
\bibinfo{author}{\bibfnamefont{S.~J.} \bibnamefont{Zweben}},
  \bibinfo{author}{\bibfnamefont{B.~D.} \bibnamefont{Scott}},
  \bibinfo{author}{\bibfnamefont{J.~L.} \bibnamefont{Terry}},
  \bibinfo{author}{\bibfnamefont{B.}~\bibnamefont{LaBombard}},
  \bibinfo{author}{\bibfnamefont{J.~W.} \bibnamefont{Hughes}},
  \bibnamefont{and} \bibinfo{author}{\bibfnamefont{D.~P.}
  \bibnamefont{Stotler}}, \bibinfo{journal}{Phys. Plasmas}
  \textbf{\bibinfo{volume}{16}}, \bibinfo{pages}{082505}
  (\bibinfo{year}{2009}).

\bibitem[{\citenamefont{Poli et~al.}(2010)\citenamefont{Poli, Bottino, Hornsby,
  Peeters, Ribeiro, Scott, and Siccino}}]{eps10}
\bibinfo{author}{\bibfnamefont{E.}~\bibnamefont{Poli}},
  \bibinfo{author}{\bibfnamefont{A.}~\bibnamefont{Bottino}},
  \bibinfo{author}{\bibfnamefont{W.~A.} \bibnamefont{Hornsby}},
  \bibinfo{author}{\bibfnamefont{A.~G.} \bibnamefont{Peeters}},
  \bibinfo{author}{\bibfnamefont{T.}~\bibnamefont{Ribeiro}},
  \bibinfo{author}{\bibfnamefont{B.~D.} \bibnamefont{Scott}}, \bibnamefont{and}
  \bibinfo{author}{\bibfnamefont{M.}~\bibnamefont{Siccino}},
  \bibinfo{journal}{Plasma Phys. Contr. Fusion} \textbf{\bibinfo{volume}{52}},
  \bibinfo{pages}{submitted} (\bibinfo{year}{2010}).

\bibitem[{\citenamefont{Naulin}(2003)}]{Naulin03}
\bibinfo{author}{\bibfnamefont{V.}~\bibnamefont{Naulin}},
  \bibinfo{journal}{Phys. Plasmas} \textbf{\bibinfo{volume}{10}},
  \bibinfo{pages}{4016} (\bibinfo{year}{2003}).

\bibitem[{\citenamefont{Naulin et~al.}(2005)\citenamefont{Naulin, Kendl,
  Garcia, Nielsen, and Rasmussen}}]{Naulin05}
\bibinfo{author}{\bibfnamefont{V.}~\bibnamefont{Naulin}},
  \bibinfo{author}{\bibfnamefont{A.}~\bibnamefont{Kendl}},
  \bibinfo{author}{\bibfnamefont{O.~E.} \bibnamefont{Garcia}},
  \bibinfo{author}{\bibfnamefont{A.~H.} \bibnamefont{Nielsen}},
  \bibnamefont{and} \bibinfo{author}{\bibfnamefont{J.~J.}
  \bibnamefont{Rasmussen}}, \bibinfo{journal}{Phys. Plasmas}
  \textbf{\bibinfo{volume}{12}}, \bibinfo{pages}{052515}
  (\bibinfo{year}{2005}).

\bibitem[{\citenamefont{Scott}(2005{\natexlab{b}})}]{zfenergy}
\bibinfo{author}{\bibfnamefont{B.}~\bibnamefont{Scott}}, \bibinfo{journal}{New
  J. Phys.} \textbf{\bibinfo{volume}{7}}, \bibinfo{pages}{92}
  (\bibinfo{year}{2005}{\natexlab{b}}).

\bibitem[{\citenamefont{Scott}(2003{\natexlab{b}})}]{transport}
\bibinfo{author}{\bibfnamefont{B.}~\bibnamefont{Scott}},
  \bibinfo{journal}{Phys. Plasmas} \textbf{\bibinfo{volume}{10}},
  \bibinfo{pages}{963} (\bibinfo{year}{2003}{\natexlab{b}}).

\bibitem[{\citenamefont{Scott et~al.}(2010)\citenamefont{Scott, Kendl, and
  Ribeiro}}]{pet09}
\bibinfo{author}{\bibfnamefont{B.}~\bibnamefont{Scott}},
  \bibinfo{author}{\bibfnamefont{A.}~\bibnamefont{Kendl}}, \bibnamefont{and}
  \bibinfo{author}{\bibfnamefont{T.}~\bibnamefont{Ribeiro}},
  \bibinfo{journal}{Contrib. Plasma Phys.} \textbf{\bibinfo{volume}{50}},
  \bibinfo{pages}{228} (\bibinfo{year}{2010}).

\bibitem[{\citenamefont{Scott}(2007{\natexlab{b}})}]{braggem}
\bibinfo{author}{\bibfnamefont{B.}~\bibnamefont{Scott}},
  \bibinfo{journal}{Phys. Plasmas} \textbf{\bibinfo{volume}{14}},
  \bibinfo{pages}{102318} (\bibinfo{year}{2007}{\natexlab{b}}).

\bibitem[{\citenamefont{Pogutse et~al.}(1998)\citenamefont{Pogutse, Smolyakov,
  and Hirose}}]{Pogutse98}
\bibinfo{author}{\bibfnamefont{I.~O.} \bibnamefont{Pogutse}},
  \bibinfo{author}{\bibfnamefont{A.~I.} \bibnamefont{Smolyakov}},
  \bibnamefont{and} \bibinfo{author}{\bibfnamefont{A.}~\bibnamefont{Hirose}},
  \bibinfo{journal}{J. Plasma Phys.} \textbf{\bibinfo{volume}{60}},
  \bibinfo{pages}{133} (\bibinfo{year}{1998}).

\bibitem[{\citenamefont{Hasegawa and Wakatani}(1983)}]{HasWak83}
\bibinfo{author}{\bibfnamefont{A.}~\bibnamefont{Hasegawa}} \bibnamefont{and}
  \bibinfo{author}{\bibfnamefont{M.}~\bibnamefont{Wakatani}},
  \bibinfo{journal}{Phys. Rev. Lett.} \textbf{\bibinfo{volume}{50}},
  \bibinfo{pages}{682} (\bibinfo{year}{1983}).

\bibitem[{\citenamefont{Strauss}(1976)}]{Strauss76}
\bibinfo{author}{\bibfnamefont{H.}~\bibnamefont{Strauss}},
  \bibinfo{journal}{Phys. Fluids} \textbf{\bibinfo{volume}{19}},
  \bibinfo{pages}{134} (\bibinfo{year}{1976}).

\bibitem[{\citenamefont{Strauss}(1977)}]{Strauss77}
\bibinfo{author}{\bibfnamefont{H.}~\bibnamefont{Strauss}},
  \bibinfo{journal}{Phys. Fluids} \textbf{\bibinfo{volume}{20}},
  \bibinfo{pages}{1354} (\bibinfo{year}{1977}).

\bibitem[{\citenamefont{Scott}(2005{\natexlab{c}})}]{bdmode}
\bibinfo{author}{\bibfnamefont{B.}~\bibnamefont{Scott}},
  \bibinfo{journal}{Phys. Plasmas} \textbf{\bibinfo{volume}{12}},
  \bibinfo{pages}{062314} (\bibinfo{year}{2005}{\natexlab{c}}).

\end{thebibliography}
\bibliographystyle{aip}
}

\end{document}